%% file: main.tex
\documentclass[acmsmall,screen]{acmart}

\setcopyright{acmlicensed}
\acmJournal{CSUR}
\acmYear{2022} \acmVolume{1} \acmNumber{1} \acmArticle{1} \acmMonth{1} \acmPrice{15.00}\acmDOI{10.1145/3565971}

\usepackage{pifont}
\usepackage{rotating}
\usepackage{tikz}
\usetikzlibrary{positioning}
\usepackage{listings}
\usepackage{booktabs}
\usepackage{multirow}
\usepackage{graphicx}
\usepackage{subcaption}

\usepackage{tcolorbox}
\newcounter{findingCounter}
\newenvironment{finding}{
	\begin{tcolorbox}[colback=blue!5!white,colframe=blue!5!white,arc=0mm,grow to left by=0mm,left=0mm,grow to right by=0mm,left=1.5mm,right=1.5mm,top=1.5mm,bottom=1.5mm]
		\textbf{Summary:}
	}
	{
	\end{tcolorbox}
}

\usepackage{tikz}

\newcommand{\code}[1]{{\ttfamily \small #1}}

\newcommand{\yes}{\ding{51}}

\usepackage{hyperref}

\begin{document}

\title{Code Search: A Survey of Techniques for Finding Code}

\author{Luca Di Grazia}
\email{luca.di-grazia@iste.uni-stuttgart.de}
\affiliation{%
	\institution{Department of Computer Science, University of Stuttgart}
	\country{Germany}
}

\author{Michael Pradel}
\email{michael@binaervarianz.de}
\affiliation{%
	\institution{Department of Computer Science, University of Stuttgart}
	\country{Germany}
}

\begin{abstract}
	\input{sections/abstract}

\end{abstract}



\maketitle

\input{sections/introduction}

\input{sections/query}

\input{sections/querypreprocessing}

\input{sections/indexingretrieval}

\input{sections/rankingpruning}

\input{sections/study}

\input{sections/future}

\input{sections/conclusion}

%\newpage
%\input{sections/random}
%\newpage

\bibliographystyle{ACM-Reference-Format}
\bibliography{references,more_references}
%more_references.bib contains papers that are not directly about code search, but they are researches that are used in the code search papers.

\end{document}

%% file: sections/abstract.tex
The immense amounts of source code provide ample challenges and opportunities during software development.
To handle the size of code bases, developers commonly search for code, e.g., when trying to find where a particular feature is implemented or when looking for code examples to reuse.
To support developers in finding relevant code, various code search engines have been proposed.
This article surveys 30 years of research on code search, giving a comprehensive overview of challenges and techniques that address them.
We discuss the kinds of queries that code search engines support, 
how to preprocess and expand queries,
different techniques for indexing and retrieving code,
and ways to rank and prune search results.
Moreover, we describe empirical studies of code search in practice.
Based on the discussion of prior work, we conclude the article with an outline of challenges and opportunities to be addressed in the future.

%% file: sections/introduction.tex
\section{Introduction}
\label{sec:intro}

%[Add a section on code corpora? (PLs, size, search in single vs. many projects, single vs multi language, etc.)]

Many kinds of information are stored in digital systems, which offer convenient access, large storage capacities, and the ability to process information automatically.
To enable people to quickly find digitally stored information, research on information retrieval has led to powerful search engines.
Today, commercial search engines are used by billions of people every day to retrieve various kinds of information~\cite{Seymour2011}, such as textual information, images, or videos.

As software is becoming increasingly important in various aspects of our lives, a particular kind of information is being produced in incredibly large amounts: source code.
A single, complex software project, such as the Linux kernel or modern browsers, easily comprises multiple millions of lines of source code.
At the popular open-source project platform GitHub, more than 60 million new projects have been created in 2020 alone~\cite{octoverseReport2020}.
The sheer amount of existing source code leads to a situation where most code to be written by a developer either has already been written elsewhere, or at least, is similar to some code that has already been written~\cite{Fuqing1999,Inoue2020, Rahman2018a}.

%As a result, code search tools are useful to save time and also money, in fact code search tools are also used to reuse existing software components.

To benefit from existing source code and to efficiently navigate complex code bases, software developers often search for code~\cite{Sadowski2015}.
For example, a developer may search through a code base she is working on to find where some functionality is implemented, to understand what a particular piece of code is doing, or to find other code locations that need to be changed while fixing a bug.
Beyond the code base a developer is working on, developers also commonly search through other projects within an organization or through open-source projects.
For example, a developer may look for examples of how to implement a specific functionality, search for usage examples of an application programming interface (API), or simply cross-check newly written code against similar existing code.
We call these and related activities \emph{code search}.
To support developers during code search, \emph{code search engines} automatically retrieve code examples relevant to a given query from one or more code bases.

At a high level, the challenges for building a successful code search engine are similar to those in general information retrieval: provide a convenient querying interface, produce results that match the given query, and do so efficiently.
Beyond these high-level similarities, code search comes with interesting additional opportunities and challenges.
As programming languages have a formally defined syntax, one can unambiguously parse source code, and then analyze and compare it based on its structural properties~\cite{aho2007compilers}.
Moreover, source code also has well-defined run-time semantics, as given by the specification of the programming language, e.g., for Java~\cite{DBLP:books/aw/GoslingJS96}, C++~\cite{DBLP:books/daglib/0007588}, or JavaScript~\cite{ecma5}.
That is, in contrast to natural language text and other kinds of information targeted by search engines, the meaning of a piece of source code is, at least in principle, well defined.
In practice, the code in a large code corpus often is written in a diverse set of programming languages, building on various frameworks and libraries, and using different coding styles and conventions~\cite{Husain2019}.
As a result, code search engines must strike a balance between precisely analyzing code in a specific language and supporting a wide range of languages~\cite{Shepherd2012}.
Finally, the language in which a query is formulated may not be the same as the language the search results are written in.
For example, many code search engines accept natural language queries or behavioral specifications of the code to retrieve, which requires some form of mapping between such queries and code~\cite{Nguyen2017}.

\begin{figure}
	\includegraphics[scale = .6]{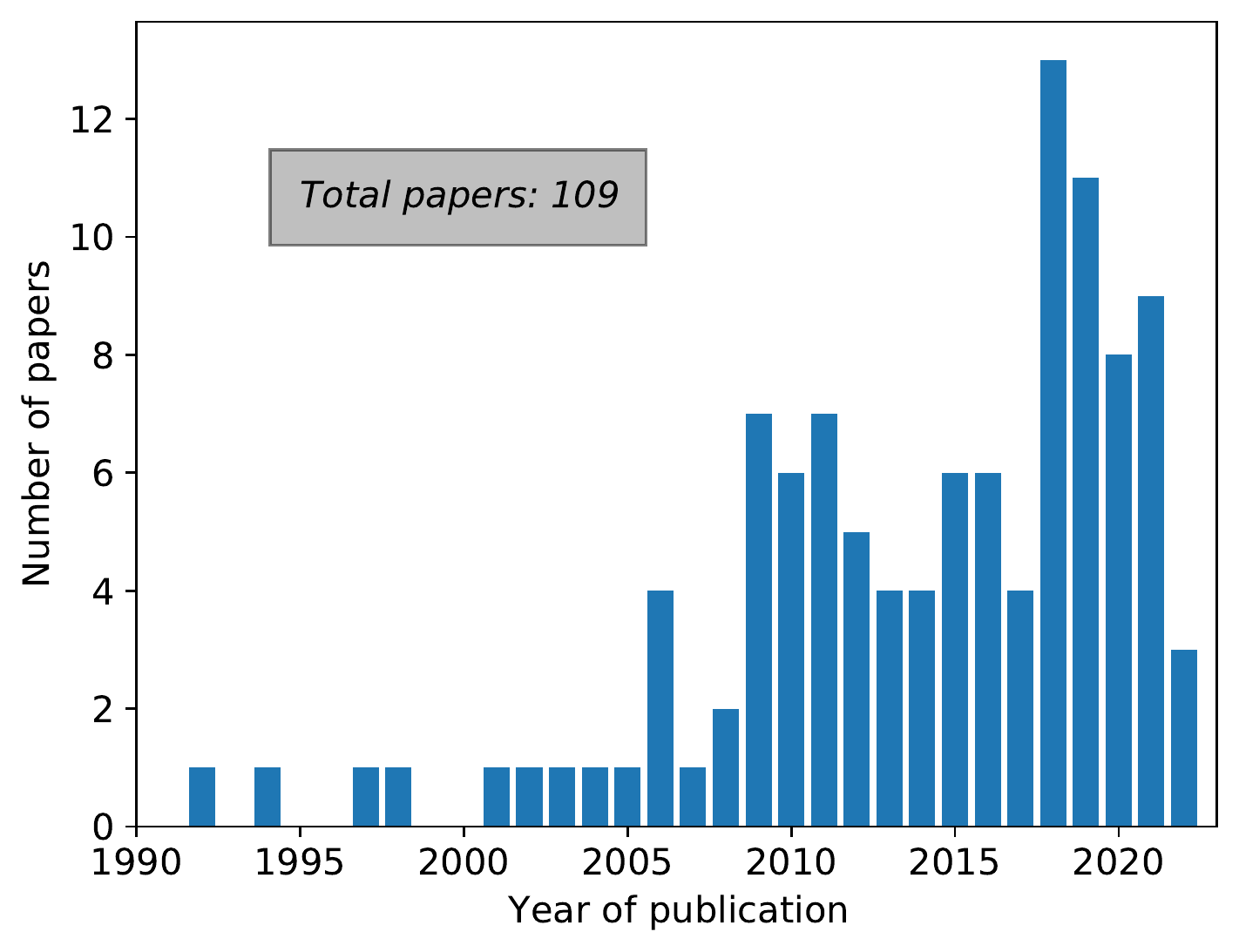}
	\caption{Papers on code search discussed in this article.}
	\label{fig:year_plot}	
\end{figure}

Motivated by the need to search through the huge amounts of available source code and by the challenges and opportunities it implies, code search has received significant attention by researchers and practitioners.
The progress made in the field is good news for developers, as they can benefit from increasingly sophisticated code search engines.
At the same time, the impressive amount of existing work makes it difficult for new researchers and interested non-experts to understand the state-of-the-art and how to improve upon it.
This article summarizes existing work on code search and describes how different approaches relate to each other.
By providing a comprehensive survey of 30 years of work on code search, we hope to provide an overview of this thriving research field.
Based on our discussion of existing work, we also point out open challenges and opportunities for future research.

\begin{figure}
	\includegraphics[width=\linewidth]{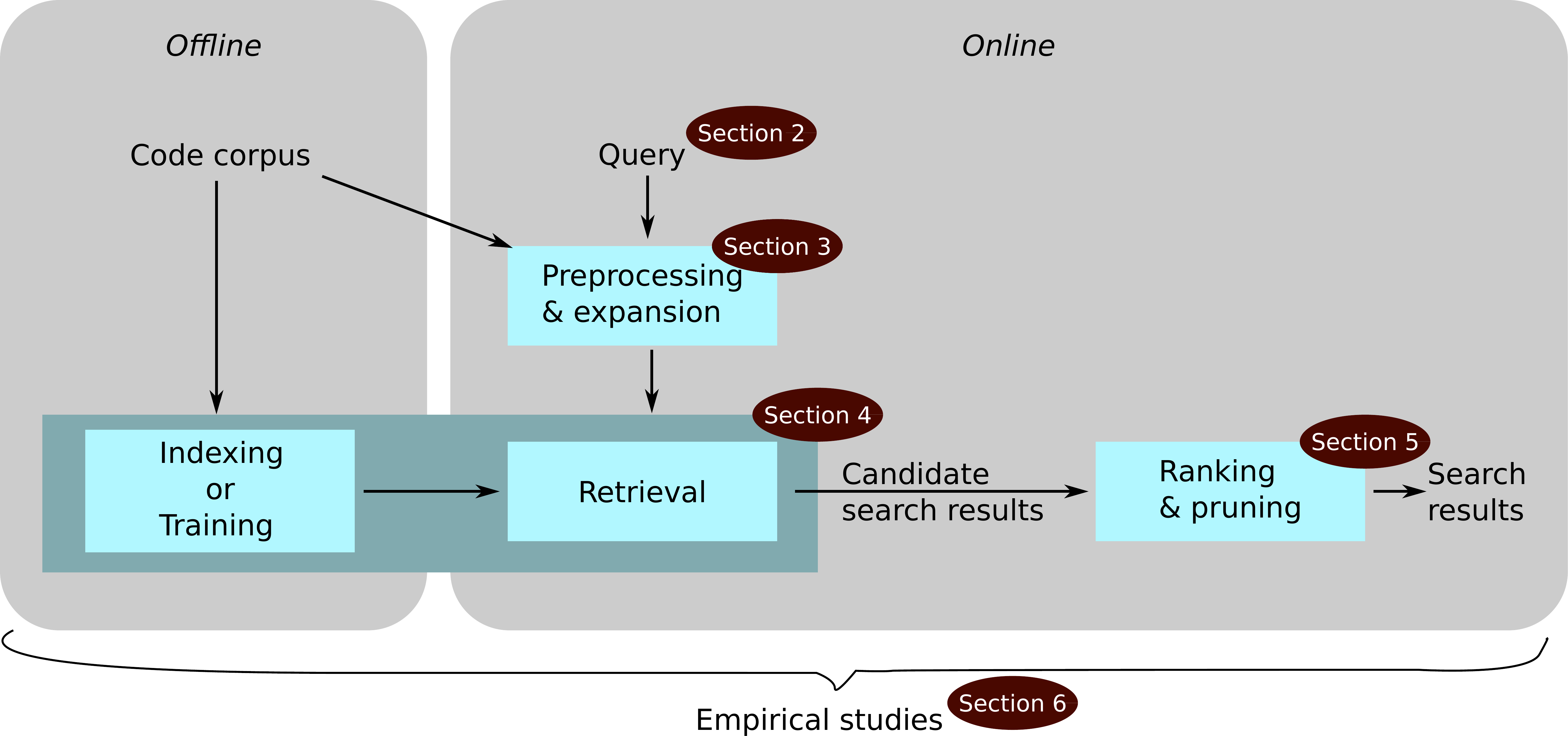}
	\caption{Overview of the topics covered in this article.}
	\label{fig:overview}
\end{figure}

Figure~\ref{fig:year_plot} shows the number of papers we discuss per year of publication, illustrating the increasing relevance of the topic.
Our survey primarily targets full research papers, i.e., more than six pages, from top-ranked conferences and journals.\footnote{Specifically, venues ranked A* or A in the CORE ranking: \href{http://portal.core.edu.au/conf-ranks}{http://portal.core.edu.au}}
In addition, we include other publications, e.g., in workshop proceedings, papers on arXiv, and technical reports, as well as publications at lower-ranked venues, if and only if they are recent (less than two years), have had a significant impact (more than ten citations), or provide a very strong match with the topic of this survey.
We use three different platforms to search for papers: Google Scholar\footnote{\href{https://scholar.google.com/}{https://scholar.google.com/}}, the ACM Digital Library\footnote{\href{https://dl.acm.org/}{https://dl.acm.org/}}, and DBLP\footnote{\url{dblp.uni-trier.de/}}.
To find an initial set of papers, we search with queries "code search" and "code retrieval".
Afterwards, we iteratively refined the set of considered papers by following citations, both backward and forward, until reaching a fixed point.
%"learning code retrieval", "api retrieval", "binary code search", "semantics code search", and "query code search".

There are several research fields related to code search that are out of the scope of this article.
In particular, we do not discuss in detail work on
general software repository mining, e.g., to extract patterns or programming rules~\cite{DBLP:journals/smr/KagdiCM07},
searching for entire applications, e.g., in an app store~\cite{Grechanik2010,McMillan2012}, and 
query-based synthesis of new code examples~\cite{DBLP:conf/oopsla/GveroK15}.
Moreover, we do not cover in detail work on code clone detection~\cite{roy2007survey} and
code completion~\cite{Bruch2009,Raychev2014}, as those are related but different problems.
Code clone detection aims at finding pieces of code that are semantically, and perhaps even syntactically, equivalent to each other, whereas code search aims at finding code that offers more details than a given query.
Code completion can be seen as a restricted variant of code search, where the code a developer has already written serves as a query to find the next few tokens or even lines to insert.
An important difference is that code search tries to retrieve existing code as-is, whereas code completion synthesizes potentially new code fragments.

Figure~\ref{fig:overview} outlines the components a typical code search engine is built from, and at the same time, gives an overview of the topics covered in this article.
Most code search engines have an offline part, which indexes a code corpus or trains a machine learning model on a code corpus, and an online part, which takes a user-provided query and retrieves code examples that match the query.
\begin{itemize}
\item Section~\ref{sec:query} presents different kinds of queries accepted by code search engines, including natural language, code snippets, formal specifications, test cases, and queries written in specifically designed querying languages.
%\item Section~\ref{sec:codepre} discusses how to preprocess and abstract the code in a corpus before it can be indexed, e.g., by parsing the code into an abstract syntax tree (AST) and by extracting particular fragments from this tree.
\item Section~\ref{sec:querypre} describes how code search engines preprocess and expand a given query, e.g., by generalizing terms in a natural language query or by lifting a given code snippet to a richer representation.
\item Section~\ref{sec:indexretrieve} presents the core component of a code search engine, which indexes code examples or trains a machine learning model, and then retrieves examples that match a query.
We discuss and compare several approaches based on how they represent the code and what kind of retrieval technique they use.
\item Section~\ref{sec:rankingpruning} presents different techniques for ranking and pruning search results before presenting them to the user, e.g., based on similarity scores between code examples and the query, or based on clustering similar search results.
\item Section~\ref{sec:studies} discusses empirical studies of developers and how they interact with code search engines, which connects the research described in the other sections to adoption in practice.
\item Section~\ref{sec:future} outlines several open challenges and research directions for future work.
\end{itemize}

Prior work surveys code search techniques from different perspectives than this article.
\citet{Garcia2006} summarize code search-related tools presented until 2006, with a focus on tools aimed at software reuse.
Another survey~\cite{Dit2013} is about techniques for locating where in a project a particular feature or functionality is implemented.
While being a problem related to code search, feature location focuses on searching through a single software project, instead of large code corpora, and on the specific use case of locating a feature, instead of the wider range of use cases covered by code search.
A short paper by \citet{Khalifa2019} discusses existing techniques for code search, focusing on information retrieval-based and deep learning-based approaches, but it covers only five papers.
Finally, another survey of code search techniques~\cite{Liu2022a} focuses on general publication trends, application scenarios where code search is used, and how search engines are evaluated.
In contrast, this article focuses more on the technical core of code search engines, including different querying languages, pre-processing of queries, ranking and pruning of results, and also empirical studies of code search in practice.

%For this survey we collect papers that can represent this active research area, describing the different techniques proposed. We selected papers from relevant software engineering conferences (A* and A) and with at least five citations. Most of them are non-learning based. Figure~\ref{fig:year_plot} shows an overview of the papers analyzed divided by year.

%% file: sections/query.tex
\section{Queries for Searching Code}
\label{sec:query}

The starting point of every search is a query.
We define a query as an explicit expression of the intent of the user of a code search engine. %, e.g. using a GUI framework~\cite{Michail2002}.
This intent can be expressed in various ways, and different code search engines support different kinds of queries.
The designers of a code search engine typically aim at several goal when deciding what kinds of queries to support:
\begin{itemize}
\item \emph{Ease}. A query should be easy to formulate, enabling users to use the code search engine without extensive training. If formulating an effective query is too difficult, users may get discouraged from using the code search engine.
\item \emph{Expressiveness}. Users should be able to formulate whatever intent they have when searching for code. If a user is unable to express a particular intent, the search engine cannot find the desired results.
\item \emph{Precision}. The queries should allow specifying the intent as unambiguously as possible. If the queries are imprecise, the search is likely to yield irrelevant results.
\end{itemize}

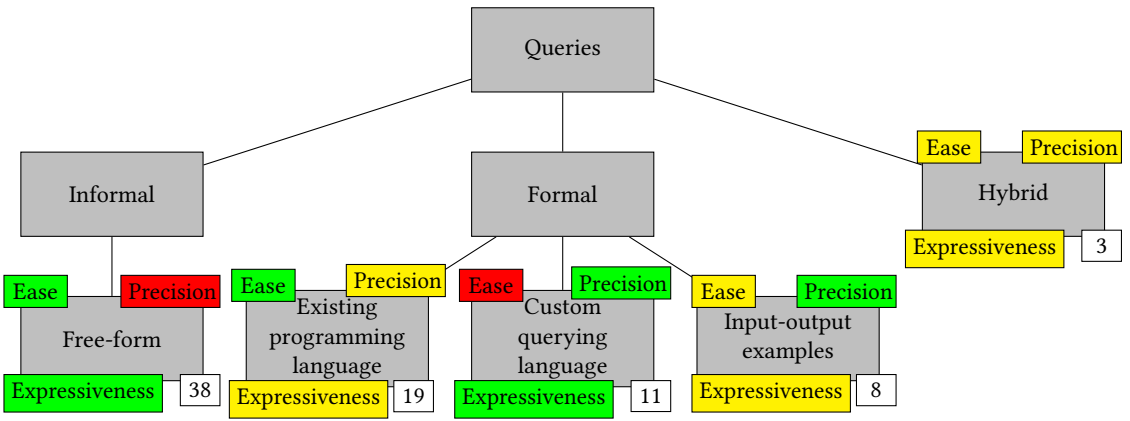
\begin{figure*}
\small
\begin{tikzpicture}[
  level distance=6em,
  every node/.style = {shape=rectangle,
    draw, align=center,
    fill=lightgray,
    text width=7em, minimum height=3.5em}
  ]
  \tikzstyle{level 1}=[sibling distance=6cm]
  \tikzstyle{level 2}=[sibling distance=3cm]
  
  \node {Queries}
    child { node (informal) {Informal}
      child { node (freeForm) {Free-form} }
    }
    child { node (formal) {Formal}
      child { node (existingPL) {Existing programming language} }
      child { node (custom) {Custom querying language} }
      child { node (io) {Input-output examples} }
    }
    child { node (semiFormal) {Hybrid}};
  
  \tikzset{every node/.style={shape=rectangle, draw, align=center, text width=1em, fill=white}}
  \node[below right=-.3em and -1em of freeForm] {38};
  \node[below right=-.3em and -1.5em of existingPL] {19};
  \node[below right=-.3em and -1em of custom] {11};
  \node[below right=-.3em and -1em of io] {8};
  \node[below right=-.3em and -1em of semiFormal] {3};
  \tikzset{every node/.style={shape=rectangle, draw, align=center, text width=2em, fill=green}}
  \node[above left=-.5em and -2em of freeForm] {Ease};
  \node[above left=-.5em and -2.1em of existingPL] {Ease};
  \tikzset{every node/.style={shape=rectangle, draw, align=center, text width=2em, fill=yellow}}
  \node[above left=-.5em and -2.5em of io] {Ease};
  \node[above left=-.5em and -2.5em of semiFormal] {Ease};
  \tikzset{every node/.style={shape=rectangle, draw, align=center, text width=2em, fill=red}}
  \node[above left=-.5em and -2.2em of custom] {Ease};
  
  \tikzset{every node/.style={shape=rectangle, draw, align=center, text width=6em, fill=red}}
  \tikzset{every node/.style={shape=rectangle, draw, align=center, text width=6em, fill=green}}
  \node[below left=-.3em and -6em of custom] {Expressiveness};
  \node[below left=-.3em and -6em of freeForm] {Expressiveness};
  \tikzset{every node/.style={shape=rectangle, draw, align=center, text width=6em, fill=yellow}}
  \node[below left=-.3em and -6em of semiFormal] {Expressiveness};
  \node[below left=-.3em and -6em of existingPL] {Expressiveness};
   \tikzset{every node/.style={shape=rectangle, draw, align=center, text width=6em, fill=yellow}}
  \node[below left=-.3em and -6.5em of io] {Expressiveness};
  
  \tikzset{every node/.style={shape=rectangle, draw, align=center, text width=3.5em, fill=red}}
  \node[above right=-.5em and -3.5em of freeForm] {Precision};
  \tikzset{every node/.style={shape=rectangle, draw, align=center, text width=3.5em, fill=green}}
  \node[above right=-.4em and -3.5em of custom] {Precision};
  \node[above right=-.5em and -3.5em of io] {Precision};
    \tikzset{every node/.style={shape=rectangle, draw, align=center, text width=3.5em, fill=yellow}}
  \node[above right=-.5em and -3.5em of semiFormal] {Precision};
  \node[above right=-.3em and -3.5em of existingPL] {Precision};
\end{tikzpicture}
\caption{Taxonomy of code search queries and number of approaches accepting each kind of query.}
\label{fig:queries taxonomy}
\end{figure*}

These goals are non-trivial to reconcile, and different code search techniques balance this trade-off in different ways.
Figure~\ref{fig:queries taxonomy} shows a taxonomy of the kinds of queries supported by existing approaches.
Broadly, we can distinguish between informal queries, formal queries, and combinations of the two.
The numbers associated with the leaf nodes of the taxonomy indicate how many papers support each kind of query.
The figure also shows how well different approaches achieve the three goals from above.
The color of the boxes containing ``Ease'', ``Precision'', and ``Expressiveness'' indicate the support for these goals, where green means strong support, yellow means medium support, and red means little support.
The remainder of this section discussed the different kinds of queries in more detail, following the structure lined out in the taxonomy.

\subsection{Free-Form Queries}
\label{sec:free-form queries}

Free-form queries are an informal way of specifying the intent of a code search.
Such a query may describe in natural language the functionality of the searched code, e.g., ``read file line by line''.
Free-form queries may also contain programming language elements, e.g., when searching for identifier names of a specific API, such as ``FileReader close''.

Free-form queries are the most commonly used kind of query in the approaches we survey~\cite{Liu2019,Chatterjee2009,Hill2009,Hill2011,Shepherd2012,Martie2015,Lu2015,Li2016,Liu2018,Sirres2018,Rahman2018,Nie2016,Michail2002,Raghothaman2016,Wang2020,Lu2018,Wu2019,Vinayakarao2017,Diamantopoulos2018,Chen2018,Sachdev2018,Linstead2009,Wang2014,Ye2020,Durao2008,Gu2018,Mcmillan2013,Chen2001,Bajracharya2010,Nguyen2017,Lv2015,Akbar2019,Ling2021,Du2021,Gu2021,Xu2021, Sun2022,Salza2022, Chai2022}
They are attractive as users can easily formulate a query, similar to using a general-purpose web search engine, with a high level of expressiveness.
On the downside, free-form queries risk being imprecise.
One reason is that natural language terms are ambiguous.
For example, the term ``float'' may refer to either a data type or to a verb.
Another reason is that the vocabulary in a query may not match the vocabulary used in a code base.
For example, a search term ``array'' may refer to a data structure that syntactically occurs as two square brackets in Java or Python~\cite{Vinayakarao2017}.

Because free-form queries are extremely versatile, different code search engines compare them against different kinds of data.
One set of approaches compares free-form queries against natural language text associated with code, e.g., API documentation~\cite{Chatterjee2009}, commit messages~\cite{Chen2001}, or words in the a project's metadata~\cite{Mcmillan2013}.
Another set of approaches compares queries against the source code, e.g., by matching the query against signatures of fields and methods~\cite{Hill2009,Hill2011} or against all identifiers in the code~\cite{Shepherd2012,Martie2015,Lu2015}.

The informal nature of free-form queries may make it difficult to accurately match a query against a code snippet, e.g., because of a vocabulary mismatch between the two.
For example, plain English queries, such as ``match regular expression'' or ``read text file''~\cite{Raghothaman2016}, may not match the terms used in the corresponding API methods.
A popular way to mitigate this mismatch is to project natural language words and source code identifiers into a common vector space~\cite{Nguyen2017} via learned word embeddings, such as Word2Vec~\cite{Mikolov2013}.
Another way to address the limitations of free-form queries is to preprocess and expand queries, which we discuss further in Section~\ref{sec:querypre}.

%Nevertheless, typing a valid query is not always easy, to help users formulate queries and provide inputs for the code search engine,~\citet{Wang2011} extend the Dependence Query Language (DQL) proposed in \cite{Wang2010query} with semantic topics.

%On the other hand, a novel search technique~\cite{Vinayakarao2017} uses information such as the position of the query word and its semantic role to calculate relevance.

%However, the feature extraction from informal queries are more challenging, but as a consequence of the increasing number of studies on Machine Learning, in the last decade researchers have improved the field of natural language processing (NLP) with significant results~\cite{Khan2016}.
%As a result, many tools accept queries in plain English with a standard tokenization is performed and some keywords are extracted~\cite{Linstead2009,Sachdev2018, Wang2014,Lu2015,Ye2020,Durao2008,Chatterjee2009,Sirres2018, Reiss2009, Gu2018,Martie2015a, Wang2020, Diamantopoulos2018,Wu2019,Chen2018,Li2016}.

% interlude to formal queries
\medskip
\noindent
To avoid the ambiguity of free-form queries and because source code is anyway written in a formal language, many code search engines support some kind of formal queries, which we discuss in the following.
The commonality of these queries is that they are written in a language with a formally specified syntax, and sometimes also formally defined semantics.

\begin{finding}
Free-form queries are easy to type and highly expressive, but they can be ambiguous and less precise than other, more formal kinds of queries.
\end{finding}

\subsection{Queries Based on Existing Programming Languages}
\label{sec:PL-based queries}

As a first kind of formal query, we start by discussing queries based on existing programming languages.
A query here is a snippet of code, possibly using some additional syntax not available in the underlying programming language.
Because developers already know the programming language they are using, such queries are easy to formulate.
The expressiveness and precision of code queries varies depending on the intent of the user and the specific search engine.

Queries based on existing programming languages roughly fall into three categories:
\begin{enumerate}
\item \emph{Plain code}. The most simple kind of code query are snippets of code as defined by the syntax of the underlying programming language~\cite{Luan2019,Balachandran2015,Fujiwara2019,Nguyen2016,Kisub2018,Takuya2011,Zhou2018,Zhou2019,Lee2010,Lee2011,Mathew2021}.
For example, the following query provides a partial implementation, for which the user seeks ways to extend it~\cite{Balachandran2015}:
\begin{lstlisting}
try {
  File file = File.createTempFile("foo", "bar");
} catch (IOException e) { }
\end{lstlisting}

\item \emph{Code with holes}. Instead of letting the search engine figure out where to extend a given code snippet, some search engines support queries that explicitly define one or more holes in the given code~\cite{Mishne2012,Mukherjee2020}.
For example, this query specifies that the user looks for how to complete the body of the given method~\cite{Mukherjee2020}:
\begin{lstlisting}
public void actionClose(JButton a, JFrame f) {
  __CODE_SEARCH__;
}
\end{lstlisting}

\item \emph{Code with pattern matching symbols}. A very precise way of describing the code to search is a query in an extension of the underlying programming language that adds patterns matching symbols.
For example, such queries may define where an expression, here denoted with \code{\#}, or a statement, here denoted with \code{@}, is missing~\cite{Paul1994,Paul1992}:
\begin{lstlisting}
if (# = #) @;
\end{lstlisting}
Such a query provides an abstract template for the code to search, and the search engine tries to retrieve some or all code snippets that the template can be extended into.
%The pattern symbols that lend the pattern language its expressive power can be classified into four broad categories: wildcards for syntactic entities, wildcards for collections of syntactic entities, named wildcards, and additional features provided to allow complex queries regarding nesting.
\end{enumerate}

A recurring challenge for search engines that accept queries written in (variants of) existing programming languages is the problem of parsing incomplete code snippets~\cite{Kisub2018, Wightman2012, Mishne2012}.
An off-the-shelf grammar of the programming language may not be able to parse a query because the query does not encompass a complete source code file or because the code is incomplete.
One way to address this problem~\cite{Kisub2018} is to heuristically fix a given code fragment, e.g., by surrounding it with additional code.

A popular kind of application of search engines that accept partial code snippets is as a source code recommendation tool.
To ensure that the recommended code matches the current context a developer is working in, e.g., the current file and project, some approaches consider the code around the actual query as context available to the search engine.
For example, \citet{Holmes2005} and \citet{Brandt2010} propose to integrate code search directly into the IDE.
%Most useful when developer knows a fragment of the code to write, e.g., the name of a specific API method, but not what kind of code is typically surrounding this fragment.
%
Other approaches~\cite{Takuya2011,Zhou2019,Mukherjee2020} spontaneously search and display example code snippets while the developer is editing a program.
The underlying idea of these approaches is that the user should not spend time on formulating the query, but simply uses the already typed code.
Finally, general code completion systems also predict code based on the existing code context while a developer is writing code.
For example, GitHub's Copilot\footnote{\url{https://copilot.github.com/}} suggests multiple lines of code using a large-scale generative neural language model~\cite{Chen2021}.
In contrast to code search, code completion synthesizes suitable code, regardless of whether exactly this code has already been written anywhere, whereas code search retrieves existing code as-is.

Instead of queries in a high-level programming language, some code search engines accept binary code as a query.
For example, an approach by \citet{David2014a} accepts a function in its compiled, binary form as a query and then searches for similar functions in a corpus of binaries.
Another approach accepts an entire binary as the query, trying to find other binaries that may be compiled from the same or similar source code~\cite{Khoo2013}.
Binary-level code search has several applications in security, e.g., to check for occurrences of known vulnerable code, and in copyright enforcement, e.g., to find code copied without permission.

%On a different scenario, an interesting outcome using this kind of queries are code clone tools. For example, the authors~\citet{Lee2010} propose a scalable instant code clone search engine for large-scale software repositories.
%The input is a source code, the dataset is clustered and the query is assigned to a cluster, then the tool checks for clones.
%\todo{Integrate \citet{Martie2017}, which is about using existing code results as a kind-of query to find similar code. May also fit in another section (I haven't read it yet).}
\begin{finding}
	Program language queries are easy to type because users do not have to learn a new language, but the expressiveness and precision of code queries varies depending on the intent of the user and the specific search engine.
\end{finding}

\subsection{Custom Querying Languages}

A common alternative to queries based on an existing programming programming language are custom querying languages.
They provide a high degree of expressiveness and precision, at the expense of reduced ease of use, because users need to learn the querying language.

\subsubsection{Logic-based Querying Languages}

The most prevalent kind of custom querying languages is first-order logic predicates that describe properties of the code to search.
For example, \citet{Janzen2003} extend the logic programming language TyRuBa\footnote{\url{http://tyruba.sourceforge.net/}} to support queries such as the following, which searches for a package with a class called ``HelloWorld''
\begin{lstlisting}
package(?P, class, ?C), class(?C, name, HelloWorld)
\end{lstlisting}
In a similar vein, \citet{Hajiyev2006} describe a code querying technique based on Datalog queries. Datalog is a logic-based language that, given a set of elements and relationships between the elements, answers queries.
The approach considers program elements, e.g., classes and methods, and several relationships between them, e.g., the fact that a class inherits from another class, or that a class has a method.
A user can query a code base by formulating logic-based queries over these elements and relationships, such as asking for all methods in a class called ``A'', where ``A'' inherits from a class called ``B''.
Other approaches support logical queries over identifiers and structural relationships between them~\cite{Sindhgatta2006,Wang2010query,Wang2011}.

Several languages allow for predicates beyond describing program elements and their relationships.
One example is to also support meta-level properties, such as how many imports a file has.
For example, the query language by \citet{Martie2015a} allows for queries such as:
\begin{lstlisting}
import count > 5 AND extends class FooBar
\end{lstlisting}
The Alice search engine~\cite{Sivaraman2019} supports a kind of semantic predicates, e.g., to search for code that calls the \code{readNextFile} method in a loop and handles an exception of a type \code{FileNotFoundException}.

\subsubsection{Significant Extensions of Existing Programming Languages}

Instead of logic-based querying languages, several search engines accept queries in custom languages that significantly extend an existing programming language.
Similar to the kinds of queries discussed in Section~\ref{sec:PL-based queries}, such queries contain fragments of an existing programming language.
One such language is by \citet{Inoue2020} who support different kinds of wildcard tokens that match any single token, any token sequence, or any token sequence discarding paired brackets, respectively.
In addition, their queries may use popular regular expression operators for choice, repetition, and grouping to enhance the expressiveness.
For example, the following query  will search for nested if-else clauses:
\begin{lstlisting}
$( if $$ else $) $+
\end{lstlisting}

Another significant extension of an existing programming language is the "semantic patch language" of Coccinelle~\cite{Lawall2018}.
It allows for describing a patch, as produced by the popular \emph{diff} tool, augmented with metavariables that match a specific piece of code and with a wildcard operator.
A query hence describes a set of rules that the old and the new code must match, which then used to search for specific code changes in the version history of a project~\cite{Lawall2016}.

\subsubsection{Other Custom Languages}

A custom querying language by \citet{Premtoon2020} describes code in a way that can be mapped to a program expression graph, which describes computations via operator nodes and dataflow edges~\cite{DBLP:conf/popl/TateSTL09}.
In contrast to the above approaches, their queries are not specific to a single programming language, but can be used to search through projects in multiple languages.

\begin{finding}
	Custom querying language queries can offer high expressiveness and precision, but are (at least initially) less easy to type because users have to learn the custom language first.
\end{finding}

\subsection{Input-Output Examples as Queries}

All kinds of queries discussed so far focus on the source code itself, but neglect an important property of code that distinguishes it from other kinds of data supported by search engines, such as text: executability.
To exploit this property, some search engines support queries that are behavioral specifications and that characterize examples of the code behavior.
Such queries typically come in the form of one or more input-output examples of the code to search.

The pioneering work by \citet{Podgurski1993} is the first to propose input-output examples as queries, and other approaches adopt and improve this idea~\cite{Reiss2009,Stolee2012,Stolee2014,Stolee2016,Jiang2018}.
For example, these search engines enable users to search for code that given the input ``susie@mail.com'' produces ``susie''~\cite{Stolee2014}.
Beyond supporting developers who search for specific kinds of code, another application of input-output-based code search is to find code fragments that can be used in automated program repair~\cite{Ke2015}.

An extended form of input-output examples are queries in the form of executable test cases~\cite{Lemos2007,Lemos2011}.
Adapting the test-driven development paradigm, the basic idea is that a developer first implements test cases for some functionality and then searches for existing code that provides the desired functionality.
Test cases here serve two purposes:
First, they define the behavior of the desired code to be searched.
Second, they test the search results for suitability in the local context.

\begin{finding}
	Using input-output examples as queries allows for precisely specifying the desired behavior, but providing sufficiently many examples to fully express this behavior may require some effort.
\end{finding}

\subsection{Hybrids of Informal and Formal Queries}

A few approaches support not only one kind of query, but hybrid queries that combine multiple of the kinds described above.
One example is the work by \citet{Reiss2009}, which in addition to input-output examples supports free-form queries. 
For example, a user may search for a method that mentions ``roman numeral'' and produces ``XVII'' for the input ``17''.
Another kind of hybrid query combines free-form, natural language terms with references to program elements~\cite{Wightman2012}, e.g., ``sort playerScores in ascending order'', where ``playerScores'' refers to a variable in the code.
Finally, \citet{Martie2017} mix free-form queries and logical queries over code properties.
For example, a query may ask for code that matches the keywords "http servlet", extends a class \code{httpservlet}, and has more than three imports.

%% file: sections/querypreprocessing.tex
\section{Preprocessing and Expansion of Queries}
\label{sec:querypre}

The query provided by a user may not be the best possible query to obtain the results a user expects.
One reason is that natural language queries suffer from the inherent imprecision of natural language.
Another reason is that the vocabulary used in a query may not match the vocabulary used in a potential search result. 
For example, a query about ``container'' is syntactically different from ``collection'', but both refer to similar concepts.
Finally, a user may initially be unsure what exactly she wants to find, which can cause the initial query to be incomplete.

%An interesting example is from \citet{Sirres2018}.
%The authors show an example in which a user asks a question about "how to get the MD5 checksum of a file in Java".
%One of the potential answers contains all the significant keywords: "Java", MD5", "checksum" etc. 
%However, the answer proposed contains also other keywords that can mismatch the quality of the search task. 

To address the limitations of user-provided queries, approaches for preprocessing and expanding queries have been developed.
We discuss these approaches by focusing on three dimensions:
(i) the user interface, i.e., if and how a user gets involved in modifying queries,
(ii) the information used to modify queries, i.e., what additional source of knowledge an approach consults, and
(iii) the actual technique used to modify queries.
Table~\ref{tab:querypre} summarizes different approaches along these three dimensions, and we discuss them in detail in the following.

\begin{table}
	\caption{Overview of approaches for preprocessing and expansion of queries.}
	\label{tab:querypre}
	\small
	\setlength{\tabcolsep}{3.3pt}
	\centering
	\begin{tabular}{@{}l|lllllllllllllll@{}}
		\toprule
		& \begin{sideways}\begin{minipage}{9em}\citet{Paul1994}\end{minipage}\end{sideways}
		& \begin{sideways}\begin{minipage}{9em}\citet{Wang2011}\end{minipage}\end{sideways}
		& \begin{sideways}\begin{minipage}{9em}\citet{Shepherd2012}\end{minipage}\end{sideways}
		& \begin{sideways}\begin{minipage}{9em}\citet{Sisman2013}\end{minipage}\end{sideways}
		& \begin{sideways}\begin{minipage}{9em}\citet{Lv2015}\end{minipage}\end{sideways}
		& \begin{sideways}\begin{minipage}{9em}\citet{Lu2015}\end{minipage}\end{sideways}
		& \begin{sideways}\begin{minipage}{9em}\citet{Martie2015a}\end{minipage}\end{sideways}
		& \begin{sideways}\begin{minipage}{9em}\citet{Li2016}\end{minipage}\end{sideways}
		& \begin{sideways}\begin{minipage}{9em}\citet{Nie2016}\end{minipage}\end{sideways}
		& \begin{sideways}\begin{minipage}{9em}\citet{Martie2017}\end{minipage}\end{sideways}
		& \begin{sideways}\begin{minipage}{9em}\citet{Sirres2018}\end{minipage}\end{sideways}
		& \begin{sideways}\begin{minipage}{9em}\citet{Rahman2018}\end{minipage}\end{sideways}
		& \begin{sideways}\begin{minipage}{9em}\citet{Lu2018}\end{minipage}\end{sideways}
		& \begin{sideways}\begin{minipage}{9em}\citet{Wu2019}\end{minipage}\end{sideways}
		& \begin{sideways}\begin{minipage}{9em}\citet{Li2019}\end{minipage}\end{sideways}
		\\
		\midrule
		User interface: \\
		\hspace{1em} Transparent to user
		&\yes&\yes&\yes&\yes&\yes&\yes&&\yes&\yes&&\yes&\yes&&\yes&\yes\\
		\hspace{1em} Based on (implicit) user feedback
		&&&&\yes&&&\yes&&&\yes&&&&&\yes \\
		
		Information used to modify queries: \\
		\hspace{1em} Initial search results
		&&&&\yes&&&\yes&&&\yes&&&&\yes \\
		\hspace{1em} Similarity of search terms and/or identifiers
		&&&&\yes&\yes&\yes&&\yes&\yes&&\yes&\yes&\yes \\
		\hspace{1em} NL/code dataset (e.g., Stack Overflow)
		&&&&&\yes&&&&\yes&&\yes&\yes \\
		\hspace{1em} Recurring code changes
		&&&&&&&&&&&&&&\yes \\
		
		Technique used to modify queries: \\
		\hspace{1em} Weigh search terms
		&&&\yes&&&&&&&&&\yes \\
		\hspace{1em} Add or replace search terms
		&&&&\yes&\yes&&\yes&\yes&\yes&&\yes&\yes&\yes&\yes \\
		\hspace{1em} Lift query to richer representation
		&\yes&\yes&&&&&\yes \\
		
		\bottomrule
	\end{tabular}
\end{table}

\subsection{User Interface of Query Preprocessing and Expansion Approaches}

\paragraph{Transparent vs. interactive}
The majority of code search engines that perform some form of query preprocessing or expansion do so in a fully transparent way, i.e., the user is not aware of this part of the approach. For example, \citet{Sisman2013} propose to automatically expand queries using similar terms from search results, while others transparently expand queries using dictionaries~\cite{Li2016} and synonymous~\cite{Lu2015}.
An exception is the work by \citet{Martie2015a,Martie2017}, where the user interactively reformulates queries based on keyword recommendations made by the search engine.
Another interactive approach~\cite{Lu2018} collects relations between words from the source code, such as that one type name inherits from another, or that a word is part of a compound word, and then removes irrelevant words using an English dictionary.
After this process the user can give a feedback about the query and iterate the query refinement until satisfied.

\paragraph{User feedback}
To improve the initially given queries, some approaches rely on feedback by the user.
% explicit: up- and down-rate results
Such feedback can be given explicitly, as in the work by \citet{Martie2015a,Martie2017}, where a user can up-rate or down-rate particular search results, which is then used to show more or less search results with similar features.
% implicit: pseudo relevance feedback, based in first results for initial query 
Instead of explicit feedback, \citet{Sisman2013} rely on so-called pseudo relevance feedback given implicitly through the highest-ranked search results retrieved for initially given query.
The approach then enriches the initial query with search terms drawn from the initial search results.
% implicit: RL based on clicked results
Another way of using implicit user feedback is by observing what search results a user clicks on, which may provide valuable information on what the user is searching for.
The Cosoch approach exploits this feedback in a reinforcement learning-based approach~\cite{Li2019}.
Their approach tracks across multiple queries which search results a user selects, and tries to maximize the normalized discounted cumulative gain (NDCG), which measures the quality of a ranked list of search results.

\begin{finding}
	User interfaces can help preprocessing queries in a transparent way or using feedback from users. 
\end{finding}

\subsection{Information Used to Modify Queries}

\paragraph{Initial search results}
Effectively modifying a query requires some information in addition to the query itself.
Several approaches use the results returned for the initially provided query for this purpose~\cite{Sisman2013,Martie2015a,Martie2017}.
A downside of relying on initial search results to modify queries is that the search must be performed multiple times before obtaining the final search results, which may negatively affect efficiency.

\paragraph{Similarity of search terms and/or identifiers}
A commonality of several approaches for query preprocessing and expansion is to compare words or identifiers in a query with those in a potential search result through some kind of similarity measure.
% positional proximity of terms
One approach~\cite{Sisman2013} builds on the observation that terms in search results that frequently appear close to terms in the query may also be relevant, and then expands the initial query with those words.
% domains-specific dictionary
Others builds on domain-specific dictionaries~\cite{Li2016} or 
% synonyms via WordNet
on synonyms~\cite{Lu2015} found using WordNet~\cite{Leacock1998} to add or replace query terms with related terms.
% learned word embeddings
A more recent alternative to curated databases of word similarities are learned word embeddings, e.g., via Word2vec~\cite{Mikolov2013}, which can help in revising queries~\cite{Rahman2018}.

\paragraph{NL/code datasets}
A third kind of information used by several approaches to revise queries are datasets of documents that combine natural language and code.
% API documentation
For example, \citet{Lv2015} use API documentation to identify which API a query is likely to refer to, and then expand the query accordingly.
Online discussion forums with programming-related questions and answers, e.g., Stack Overflow also have been found to help in revising queries~\cite{Nie2016,Rahman2018,Sirres2018}.
These approaches search online posts related to a given query, and then extract additional relevant words, software-specific terms, and API identifiers to augment the query.
Since the questions and answers cover various application domains and are curated based on feedback by thousands of developers, they provide a valuable dataset to associate natural language words with related programming terms.

\paragraph{Recurring code changes}
Motivated by the observation that developers may have to adapt a retrieved code example, e.g., to use the most recent version of an API, \citet{Wu2019} expand queries to proactively consider such potential code adaptations.
At first, their approach mines recurring code changes from version histories of open-source projects, which provides information such as that a code token \code{A} is often changed to a code token \code{B}.
Given a user query, they then retrieve matching code examples, and if these examples include a frequently changed token, say \code{A}, expand the query with the updated token, say \code{B}.
With the expanded query, the search engine hence will retrieve updated versions of the code examples, freeing the developer from adapting the code manually.

\begin{finding}
	To automatically modify queries, code search engines most commonly use similarities between search terms and identifiers, as well as corpora of natural language and code.
\end{finding}

\subsection{Techniques Used to Modify Queries}

\paragraph{Weigh search terms}
The perhaps most straightforward way of augmenting a given search query is to weigh the given search terms.
Several code search engines implement this idea~\cite{Shepherd2012,Rahman2018}, with the goal of giving terms that are most relevant for finding suitable results at a higher weight.
For example, \citet{Rahman2018} estimate the weights of API class names by applying the page rank algorithm~\cite{Brin1998} to an API co-occurrence graph.

\paragraph{Add or replace search terms}
% similarity between search terms and APIs by checking for identifier in method name and body
Another common technique is to add or replace terms in the given search query, e.g., by adding terms that are related or synonymous to those already in the query.
\citet{Lv2015} propose a query refinement technique specifically aimed at APIs.
In a two-step approach, they first identify an API that the query is likely to refer to, and then expand the original query with identifier names related to this API.
For example, for an initial query ``how to save an image in png format'', the approach may identify the API method \code{System.Drawing.Image.Save} to be likely relevant, and hence, adds the fully qualified method name into the search query.

\paragraph{Lift query to richer representation}
To ease matching a query against potential search results, several approaches lift the query into a richer representation.
An early example is the SCRUPLE tool by \citet{Paul1994}.
It transforms the query specified by the user with a pattern parser into an extended nondeterministic finite automaton called code pattern automaton. 
Other approaches lift queries into a graph representation.
For example, \citet{Wang2011} take a query formulated in a custom querying language and then transforms it into a graph representation that expresses call relations, control flow relations, and data flow relations.
As another example, \citet{Li2016} transform a natural language query into an ``action relationship graph'', which expresses sequencing, condition, and callback relationships between parts of the code described in the query.
For example, given a query ``add class 'checked' to element and fade in the element'', their approach would infer that the two parts combined by ``and'' are supposed to happen in sequence.
%Both approaches~\cite{Wang2011,Li2016} exploit the graph representation of a query to view the search problem as a graph matching problem.

\begin{finding}
	The most popular techniques to modify queries are using weighing, adding, or replacing search terms, as well as lifting queries to a richer representation.
\end{finding}

%% file: sections/indexingretrieval.tex
\section{Indexing or Training, Followed by Retrieval of Code}
\label{sec:indexretrieve}

The perhaps most important component of a code search engine is about retrieving code examples relevant for a given query.
The vast majority of approaches follows a two-step approach inspired by general information retrieval:
At first, they either index the data to search through, e.g., by representing features of code examples in a numerical vector, or train a model that learns representations of the data to search through.
Then, they retrieve relevant data items based on the pre-computed index or the trained model.
To simplify the presentation, we refer to the first phase as ``indexing'' and mean both indexing in the sense of information retrieval and training a model on the data to search through.

The primary goal of indexing and retrieval is effectiveness, i.e., the ability to find the ``right'' code examples for a query.
To effectively identify these code examples, various ways of representing code and queries to compare them with each other have been proposed.
A secondary goal, which is often at odds with achieving effectiveness, is efficiency.
As users typically expect code search engines to respond within seconds~\cite{Sadowski2015}, building an index that is fast to query is crucial.
Moreover, as the code corpora to search through are continuously increasing in size, the scalability of both indexing and retrieval is important as well~\cite{Babenko2016}.

\begin{figure}
\includegraphics[width=\linewidth]{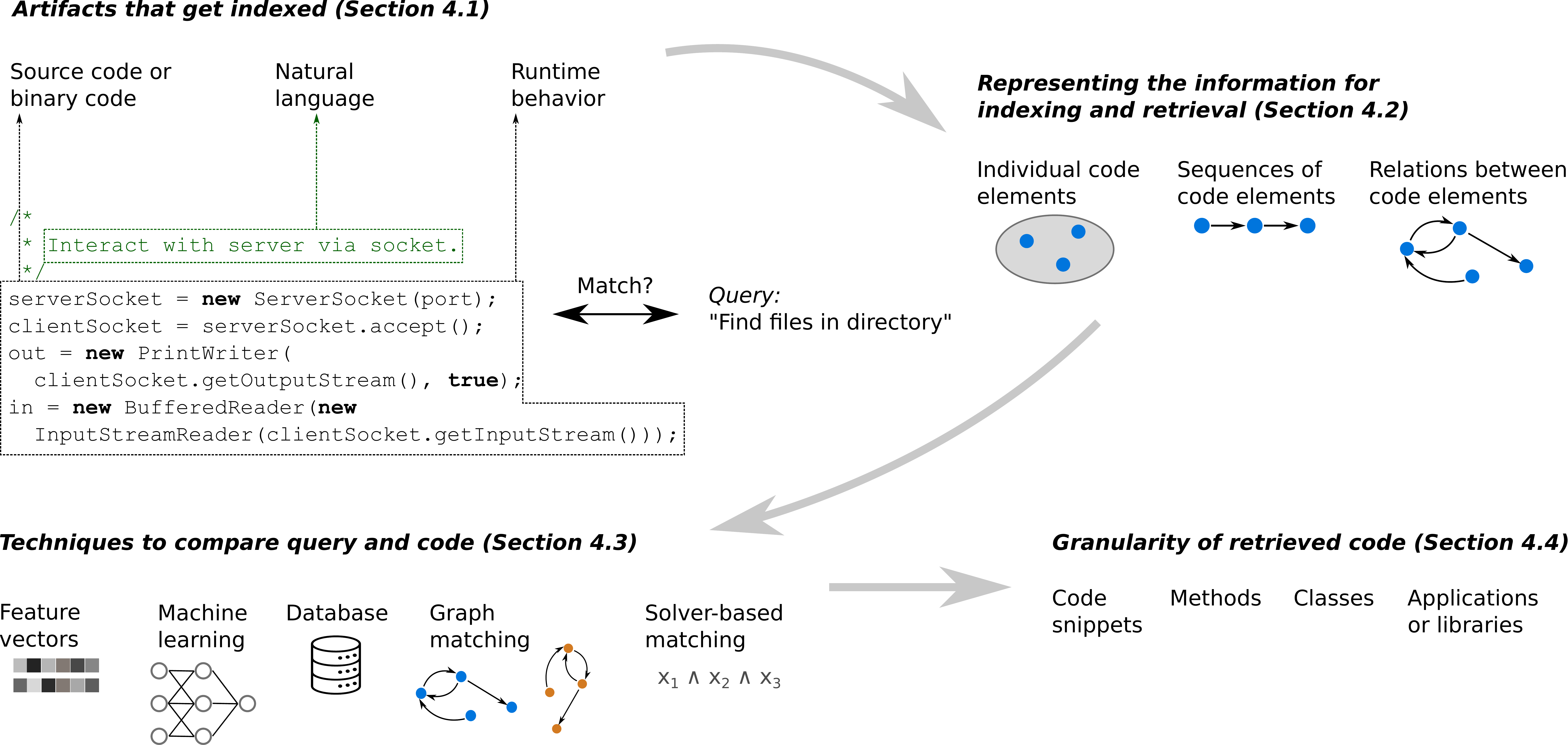}
\caption{Overview of techniques for indexing and retrieval.}
\label{fig:irOverview}
\end{figure}

We survey the many different approaches to indexing, training and retrieval in code search engines along four dimensions, as illustrated in Figure~\ref{fig:irOverview}.
Section~\ref{sec:ir artifacts} discuss what kind of artifacts a search engine indexes.
Section~\ref{sec:ir kinds} describes different ways of representing the extracted information.
Section~\ref{sec:ir techniques} presents techniques for comparing queries and code examples with each other. 
Table~\ref{tab:IR_and_information} summarizes the approaches along these first three dimensions.
Finally, Section~\ref{sec:ir granularity} discusses different levels of granularity of the source code retrieved by search engines.

\begin{table}[]
	\caption{Overview of approaches based on information retrieval technique respect to kind of indexed information (1-3 rows) and kind of feature extracted (4-6) rows.}
	\label{tab:IR_and_information}
	\centering
	\small
	\begin{tabular}{@{}lp{9em}p{7em}p{7em}p{7em}@{}}
		\toprule
		& \multicolumn{3}{c}{IR technique} \\
		\cmidrule{2-4}
		&  Feature vectors & Machine learning & Other \\ \midrule
		Indexed artifact:\\
		\hspace{1.5em}Source code  
		& \cite{Zhou2019,Lee2011,Balachandran2015,Diamantopoulos2018,Lee2010,Luan2019,Khoo2013,Shepherd2012,Fujiwara2019,Wu2019,Bajracharya2009,Nguyen2017,Nie2016,Sirres2018} 
		& \cite{Husain2019,Sisman2013,Sivaraman2019, Salza2022}
		& \cite{Li2016,Paul1992,Paul1994,David2014a,Hajiyev2006,Martie2015a,Inoue2020,Holmes2005}\\
		
		\hspace{1.5em}Runtime behavior &   
		\cite{Podgurski1993}  & 
		\cite{Jiang2018,Mishne2012,Stolee2016,Poshyvanyk2009} \\
		
		\hspace{1.5em}Natural language  &    
		\cite{Chen2018,Cambronero2019,Takuya2011,Mcmillan2013,Bajracharya2010,Kisub2018} & 
		\cite{Sachdev2018,Gu2018,Linstead2009,Mukherjee2020, Sun2022, Chai2022} & 
		\cite{Ossher2009,Chen2001} \\
		
		\midrule
		
		Representation of indexed code:\\
		
		\hspace{1.5em}Individual code elements &   
		\cite{Diamantopoulos2018,Khoo2013}  & 
		\cite{Chen2018, Salza2022, Chai2022} & 
		\cite{Inoue2020,Chen2001} \\
		
		\hspace{1.5em}Sequences of code elements &  
		\cite{Wu2019,Takuya2011,Wu2019,Bajracharya2010, Kisub2018}  & 			
		\cite{Gu2018} & 
		\cite{David2014a} \\
		
		\hspace{1.5em}Relationships between code elements
		& \cite{Lee2011,Balachandran2015, Luan2019, Lee2010, Nie2016, Lv2015} 
		& \cite{Linstead2009,Mishne2012,Nguyen2016, Sun2022}
		& \cite{Hajiyev2006, Paul1992, Paul1994,Beckmann1990,Li2016, Holmes2005}\\
		\bottomrule
	\end{tabular}%
\end{table}
%
%\todo{figure to illustrate the main approaches?}

\subsection{Artifacts That Get Indexed}
\label{sec:ir artifacts}

When creating an index of code examples to retrieve, code search engines consider different artifacts related to code.

\subsubsection{Source Code and Binary Code}

The most obvious, and by far most prevalent, artifact to index is the source code itself.
Many approaches target a high level programming language, such as Java~\cite{Zhou2019,Balachandran2015,Diamantopoulos2018,Lee2010,Bajracharya2006,Fujiwara2019,Holmes2005,Sivaraman2019,Nguyen2017,Hajiyev2006,Xu2021}, JavaScript~\cite{Li2016,Lee2011}, and C~\cite{Paul1994, Paul1992}.
Some search engines support not only one but multiple languages~\cite{Husain2019, Salza2022, Chai2022}, e.g., Sando~\cite{Shepherd2012} (C, C++, and C\#), ccgrep~\cite{Inoue2020} (C, C++, Java, and Python), Aroma~\cite{Luan2019} (Hack, Java, JavaScript, and Python), DGMS~\cite{Ling2021} (Java and Python), and COSAL~\cite{Mathew2021} (Java and Python). 
%
%\emph{APIs.} Some works help users to find relevant APIs taken from popular libraries~\citet{Nguyen2016}. Most of the approaches search for Java APIs
%\citet{Martie2015a,Chatterjee2009,Sirres2018,Bajracharya2009} and C\# APIs~\citet{Wu2019}.
%
Instead of source code, some approaches focus on compiled code~\cite{Khoo2013,David2014a}, which is useful, e.g., to find functions in binaries that are similar to known vulnerable functions.
These approaches first disassemble a given binary and then index disassembled functions or entire binaries. 

\subsubsection{Runtime Behavior of Code}
\label{sec:runtime behavior}

Instead of only statically analyzing and indexing code, some search engines exploit the fact that source code can be executed by analyzing the runtime behavior of the code to search through.
Considering runtime behavior may be useful, e.g., when two snippets of code have similar source code but nevertheless perform different behavior.
The first code search engine that considers runtime behavior is by \citet{Podgurski1993}.
Their approach expects the user to provide inputs to the code to find, and then searches for suitable code examples by sampling the behavior of candidate code examples.
\citet{Reiss2009} select candidates using keywords and then they apply different kinds of transformations to have solutions with different behavior. To validate the dynamic behavior of the candidates with respect to the requirements given by the user, they run a set of test suites. 
Another line of work symbolically executes code to gather constraints that summarize the runtime behavior~\cite{Stolee2016,Jiang2018}.
Finally, the COSAL search engine~\cite{Mathew2021} compares the behavior of code snippets based on an existing technique for clustering code based on its input-output behavior~\cite{Mathew2020}. 

%Finally, search engines specialized on API retrieval may analyze the runtime behavior of APIs.
%Moreover, \citet{Mishne2012} capture the behavior of the program as the sequence of API calls.
%The basic idea is that the same sequence of API calls has the same behavior.

\subsubsection{Natural Language Information Associated with Code}

Beyond the source code and its runtime behavior, another valuable artifact is natural language information associated with code.
For example, such information comes in the form of comments, API documentation, commit messages, and discussions in online question-answer forums.
Several code search engines leverage this information, in particular approaches that retrieve code based on natural language queries, e.g., after training neural models on pairs of natural language descriptions and source code~\cite{Sun2022, Chai2022}.

One group of approaches leverages regular comments and structured comments that provide API documentation, e.g., by considering comments as keywords to compare a query against~\cite{Linstead2009} or by mapping code and natural language into a joint vector space~\cite{Gu2018,Sachdev2018}.
Another direction is to consider commit messages in a version control system, based on the assumption that the words in a commit message describe the source code lines affected by the commit~\cite{Chen2001}.
Finally, online discussion forums, such as Stack Overflow,\footnote{\url{https://stackoverflow.com/}} provide a dataset of pairs of code snippets and natural language descriptions, which several code search engines use to associate natural language words and code~\cite{Nie2016,Chen2018,Cambronero2019}.

\begin{finding}
	The by far most common kind of artifact that gets indexed are source code and binary code. However, there also are search engines that index traces of runtime behavior and natural language information associated with code. 
\end{finding}

\subsection{Representing the Information for Indexing and Retrieval}
\label{sec:ir kinds}

After discussing what artifacts different approaches extract information from, we now consider how this information is represented for indexing and retrieval.
We identify three groups of approaches, presented in the following with increasing levels of complexity: representations based on individual code elements, on sequences of code elements, and on relations between code elements.

\subsubsection{Individual Code Elements}

The first group of approaches focuses on individual code elements, e.g., tokens or function calls, ignoring their order and any other kind of relationship they may be in~\cite{Inoue2020,Chen2001}.
To index the code examples, these approaches then represent a code snippet as a set of code elements.
One example is work that represents a code example as a bag of tokens, and a natural language query as a bag of words~\cite{Takuya2011,Chen2018}.
Another approach represents binaries as a set of tokens extracted from disassembled binaries~\cite{Khoo2013}.
Finally, \citet{Diamantopoulos2018} represent API usages as a set of API calls, replacing each method by its type signature.
The main benefit of indexing sets of individual code elements is the conceptual simplicity of the approach, which facilitates instantiating an idea for a particular target language.
On the downside, the order of code elements and other kinds of relationships may provide useful information for precisely matching a code example against a query.

\subsubsection{Sequences of Code Elements}

To preserve ordering information of code elements during the indexing, several approaches extract sequences of code elements from a given code example.
Most commonly, these sequences are extracted in an AST-based, static analysis that focuses on particular kinds of nodes.
For example, \citet{Gu2018} represent API usages by extracting sequences of API calls from an AST. 
Another example is FaCoY~\cite{Kisub2018}, which represents a code example as a sequence of tokens extracted from an AST, where each token comes with a token type, e.g., method call or string literal.
To represent incomplete code examples, they insert empty statements to complete the snippet. \citet{David2014a} instead use control flow graphs to represent information for the binary source code.
\citet{Sun2022} represents code as a sequence of low-level instructions, which the approach obtains by compiling and then disassembling the code.
Finally, deep learning-based code search approaches often tokenize source code using a sub-word tokenizer, such as the WordPiece~\cite{Wu2016a} tokenizer used, e.g., by \citet{Salza2022}.
%In a similar learning approach,~\citet{Chai2022}, during the pre-training phase, replace 15\% random tokens with a special token [MASK] and then they let the neural model predict the token.

\subsubsection{Relations between Code Elements}

Going beyond individual code elements and sequences thereof, many approaches extract a richer set of relations between code elements.
The most common approach is to focus on entities, typically code elements, such as classes, methods, and statements, and relations between them, such as one class inheriting from another class, one method calling another method, and a method containing a statement~\cite{Balachandran2015}.
Popular examples of this approach include CodeQuest~\cite{Hajiyev2006} and Sourcerer~\cite{Linstead2009}, which extract code elements and their relations through an AST-based analysis.
Sourcerer also serves as the basis for other code search approaches, e.g., by \citet{Bajracharya2010} and \citet{Lv2015}.

\citet{Sirres2018} extract structural code entities of Java source files, collecting the relationship of imports, classes, methods and variables.
A more recent example is Aroma~\cite{Luan2019}, which parses code into a simplified parse tree and then extracts different kinds of features based on the tokens in the code, parent-child relationships, sibling relationships, and variable usages, focusing. 
In a similar vein, \citet{Ling2021} represent code as a graph that includes structural parent-child relationships, next-token relationships, and definition-use information.
\citet{Li2016} extract from ASTs three kinds of relationships between code elements: sequencing methods calls, callback between methods and methods as conditions of if statements. \citet{Holmes2005} use heuristics to collect relationships of methods inheritance, methods calls and methods usage.
Finally, Paul et al.~\cite{Paul1994, Paul1992} use non-deterministic finite automata, called code pattern automata, to represent relationships between code elements.

%\citet{Balachandran2015} generate ASTs using Eclipse JDT core parser.
%They replace the nodes with variable names with types.
%However, they keep information about the types, identifier names, operators, and string literals.

Instead of a relatively lightweight static extraction of information to index, some search engines rely on more sophisticated static analysis.
For example, \citet{Mishne2012} propose a static type state analysis that extracts temporal specifications in the form of deterministic finite-state automata that capture sequences of API method calls.
Another example is work by \citet{Premtoon2020}, which represent code examples as data flow graphs.

\begin{finding}
	To index source code examples, code search engines typically represent the code as sets of individual code elements, sequences of code elements, or as relationships between code elements.
\end{finding}

\subsection{Techniques to Compare Queries and Code}

\label{sec:ir techniques}

%\begin{figure}
%	\begin{subfigure}[t]{.497\linewidth}
%		\includegraphics[width=\linewidth]{img/s4_vector}
%		\caption{Feature Vectors and Distance Metrics.}
%	\end{subfigure}
%	\begin{subfigure}[t]{.497\linewidth}	\includegraphics[width=\linewidth]{img/s4_boolean}
%		\caption{Set-based Comparison.}
%		%	\label{fig:astsCodeChange}
%	\end{subfigure}
%	%	\hspace{1em}
%	%\vspace{.5em}
%	
%	\caption{Illustration example of the approaches with individual token representation.\todo{how code is mapped in the feature vectors. different example. }}
%	\label{fig:illustration_models}
%\end{figure}

After extracting the information from source code, execution behavior, and natural language information associated with the code, most search engines index the extracted information to then quickly respond to queries based on the pre-computed index.
The following discusses different approaches for comparing queries and code, which we group into techniques based on feature vectors computed without machine learning (Section~\ref{sec: ir features}), machine learning-based techniques (Section~\ref{sec:ir learning}), database-based techniques (Section~\ref{sec:ir database}), graph-based matching (Section~\ref{sec:ir graphs}), and solver-based matching (Section~\ref{sec:ir solver}).
%The following discusses some representative approaches; Table~\ref{tab:ir techniques} categorizes all surveyed papers into these four groups.
%
%\begin{table}
%\caption{Techniques to compare queries and code.\todo{update and complete}}
%\label{tab:ir techniques}
%\begin{tabular}{@{}p{9em}p{9em}p{9em}p{9em}@{}}
%\toprule
%Feature vectors & Machine learning & Database & Other \\
%\midrule
%\cite{Poshyvanyk2009,Bajracharya2009,Bajracharya2010,Khoo2013,Martie2015a,Sirres2018,Kisub2018} .. & ...\\
%\bottomrule
%\end{tabular}
%\end{table}

\subsubsection{Indexing and Retrieval Based on Algorithmically Extracted Feature Vectors}
\label{sec: ir features}

Several techniques are based on feature vectors and distances between these vectors. In this sub-section we discuss approaches that compute feature vectors algorithmically, i.e., without any machine learning model. 
Their general idea is to represent both the code examples and the query as feature vectors, and to then retrieve code examples with a vector similar to that of the query.
Because performing a pairwise comparison of the query vector with each code vector is inefficient, the approaches compute an index into the feature space that allows them to efficiently retrieve a ranked list of vectors similar to a given vector.

There are different ways of mapping information about code examples and queries into feature vectors.
One approach is boolean vectors~\cite{Salton1983} that express whether some feature, e.g., a particular type of AST node, are present~\cite{Luan2019}.
Another common approach is to map a set of tokens or words into a term frequency-inverse document frequency (TF-IDF) vector, which expresses not only whether a feature is present, but also how important its presence is in comparison with other features~\cite{Takuya2011,Sisman2013,Diamantopoulos2018,Wu2019}.

A popular implementation of feature-based indexing and retrieval is the Lucene library.\footnote{\url{https://lucene.apache.org/}}
Originally designed for text search, Lucene is used in various code search engines~\cite{Poshyvanyk2009,Bajracharya2009,Bajracharya2010,Khoo2013,Martie2015a,Sirres2018,Kisub2018}.
It combines the boolean model, which removes candidate vectors that do not provide the required features, and the vector space model, which computes a distance between the remaining candidate vectors and the query vector.
The feature vectors are based on a custom term-frequency formula.\footnote{https://lucene.apache.org/core/3\_5\_0/scoring.html} Moreover, \citet{Nguyen2017} use a revised Revised Vector Space Model (rVSM). The rVSM splits each token in separate words and computes the weight for each word using TF-IDF.

Instead of building upon an existing indexing and retrieval component, some search engines implement their own indexing and retrieval technique.
For example, \citet{Lee2010} use R*trees~\cite{Beckmann1990}, which recursively partition the code examples into a tree structure that can then be used to efficiently find the nearest neighbors of a query.
\citet{Luan2019} identify those code examples that have the most overlap with the query vector by representing the feature set as a sparse vector and by then computing the overlap between queries and code examples via matrix multiplication.
Another approach~\cite{Balachandran2015} matches a code query against code examples based on feature vectors for different AST subtrees of the code examples, pruning the large number of combinations to compare by considering only subtrees with the same parent node type.

\subsubsection{Learning-based Retrieval}
\label{sec:ir learning}

Neural software analysis~\cite{NeuralSoftwareAnalysis} is becoming increasingly popular, and neural information retrieval~\cite{Mitra2018} offers an attractive alternative to more traditional techniques.
Most work takes an end-to-end neural learning approach, where a model learns to embed both queries and code examples into a joint vector space.
Given this embedding, code search reduces to finding those code examples that are the nearest neighbors of a given query.
We discuss approaches following this overall pattern in the following, focusing at first on natural language-to-code search and then on code-to-code search.

%% end-to-end neural learning; map query and code in same space; retrieval=NN search

\paragraph{Learning-based natural language-to-code search}
\citet{Gu2018} pioneered with the first neural, end-to-end, natural language-to-code search engine.
Their model embeds the code of methods using three submodels that apply recurrent neural networks to the name of the method, the API sequences in the method, and all tokens in the method body, respectively.
Likewise, the model embeds the words in the query using another recurrent neural network.
All embedding models are trained jointly to reduce the distance of matching code-query pairs while keeping unrelated pairs apart.
In a similar way, \citet{Sun2022} embed a code example and a natural language description into a joint vector space.
They improve upon earlier work by translating the code into a natural language-like representation based on transformation rules.
%use two encoders: one for the source code and one for the code comment using the same shared word mapping function. The query in natural language from the user is processed using the same encoder used for the code comments and the generated embeddings are compared with the embedding in the training set.
\citet{Chen2018} use two jointly trained auto encoders to map code and text into a vector space, respectively.
\citet{Cambronero2019} compare different ways to implement neural code search, including unsupervised~\cite{Sachdev2018} and supervised approaches and different neural models~\cite{Gu2018, Husain2018}.
Because a single model may not capture all aspects of a code example, \citet{Du2021} propose an ensemble model that combines three neural code encoders, which focus on the structure of code, its variables, and its API usages, respectively. 

To foster further comparisons, the CodeSearchNet challenge~\cite{Husain2019} offers a dataset of  2 million pairs of code and natural language queries, along with several neural baseline models and ElasticSearch.\footnote{\url{https://www.elastic.co/elasticsearch/}}
Improvements on learning vector representations of code further improve the effectiveness of learning-based code search.
For example, learn from multiple code representations~\cite{Gu2021}, apply attention-based neural networks~\cite{Xu2021}, or learn from a graph representation of code and queries via a graph neural network~\cite{Ling2021}.

\paragraph{Learning-based code-to-code search}
To find code based on an incomplete code example, several learning-based approaches have been proposed.
One approach expands an incomplete code snippet using an LSTM-based language model and then searches for similar code snippets via a scalable clone detection technique~\cite{Zhou2018}.
An improved version of the approach~\cite{Zhou2019} uses a library-sensitive language model for expanding the given code snippet.
Another approach for retrieving code given an incomplete code snippet learns a model that predicts the probability that a complete code example fits the given snippet~\cite{Mukherjee2020}.
The model is based on various kinds of contextual information, e.g., the types, API calls, and code comments found around the given code snippet.

%% MP: removing the paper; seems incomprehensible and is published at a C-level venue
%\citet{Fujiwara2019} generate training data for code-to-code search by .. from the target projects and then it trains a neural model with this data. This approach is divided into four steps. First, the authors generate positive data from these code block sets. Next, they extract code blocks from projects which are not included in the target projects, they generate negative data from these code blocks. The second step generates similar code blocks as data augmentation techniques. The third step generates feature vectors based on reserved keywords and identifiers extracted from positive data for training. Last, the neural model is trained.

%% based on pre-trained word/token embeddings

\paragraph{Search based on pre-trained models}
Recent approaches use large pre-trained language models~\cite{Feng2020,Guo2021}, such as BERT~\cite{Devlin2018}, for code search.
For example, \citet{Salza2022} pre-train a BERT model on multiple programming languages  and then they fine-tune the model using two encoders: one for natural language queries and another for code snippets.
\citet{Chai2022} show the value of transfer learning for code search by pre-training CodeBERT~\cite{Feng2020} on Java and Python, applying a meta-learning approach called MAML (Model-Agnostic Meta-Learning)~\cite{Chelsea2017} to adapt the neural model to the target language, and finally fine-tuning the model with a dataset from the target language.
%
%\citet{Zhou2019} fine-tune BERT with a manually labeled dataset.
%
Instead of an end-to-end neural search that maps entire code examples and queries into a joint vector space, one can also use pre-trained embeddings of individual words and tokens.
For example, \citet{Ling2021} use GloVe~\cite{Pennington2014} and \citet{Zhou2018} use pre-trained FastText embeddings.\footnote{\url{https://fasttext.cc}}
\citet{Sachdev2018} propose an approach that maps individual code tokens into vectors, then computes a TFIDF-weighted average of them, and finally uses the resulting vector for a nearest neighbor-based search in the vector space.

%Figure~\ref{fig:learning_model} shows an example, where a pre-trained neural models gets a natural language query and it returns a code recommendation.
%
%\begin{figure}
%	\begin{subfigure}[t]{.497\linewidth}
%		\includegraphics[width=\linewidth]{img/s4_prob}
%	\end{subfigure}
%	%	\hspace{1em}
%	%\vspace{.5em}
%	
%	\caption{Illustration example of a learning approach.}
%	\label{fig:learning_model}
%\end{figure}

%% non-neural, probabilistic models
%% MP: removed the following because about code completion rather than code search
%recommend API calls
%learn from code changes: given code context, predict most likely call to insert
%based on frequencies of co-occurrences
%\citet{Nguyen2016} analyze the contributions of each candidate to compute the likelihood.
%The model takes in consideration also the context and the scope of the changes.
%The function for the score is: \emph{score(c, (C,T))}, where \emph{c} is the code change, \emph{C} is the change context, and \emph{T} is the code context.

\subsubsection{Database-based Indexing and Retrieval}
\label{sec:ir database}

Given the success of databases for storing and retrieving information, several code search approaches build upon general-purpose databases.
David et al.~\cite{David2014a} describe a code search engine for binaries that stores short execution traces (``tracelets'') in the NoSQL database MongoDB.
Given a function as a query, the approach then retrieves other functions by querying the database for matching tracelets.
Another database-based approach is by \citet{Hajiyev2006}, who build upon a relational database.
Their approach stores facts extracted from a program, such as return relationships, method calls, and read and write fields, and then formulates search queries as database queries.
In contrast to the similarity-based retrieval techniques discussed above, databases retrieve code examples that precisely match a query.

\subsubsection{Graph-based Indexing and Retrieval}
\label{sec:ir graphs}

Given a graph representation of queries and code, another common approach is to retrieve code via graph-based matching.
%
% graph similarity
\citet{Li2016} abstract both code snippets and a natural language query into graphs that represent different API method calls and their relationships.
Then, they address the retrieval problem as a search for similar graphs.
%
% rewrite and match
The Yogo search engine~\cite{Premtoon2020} represents a given query code example and all code examples to search through as dataflow graphs.
To match queries with code examples, the approach then applies a set of rewrite rules to check if the rewritten graphs match.

Instead of matching graphs, another direction is to use a graph representation of code to compute a similarity score.
% propagation through call graph
\citet{Mcmillan2013}'s Portfolio technique first computes the pairwise similarity of a query and a set of functions, and then propagates the similarity score using the spreading activation algorithm through a pre-computed call graph.
In an orthogonal step, the approach also computes the importance of every function by applying the page rank algorithm to the call graph.
Finally, the two scores are combined to retrieve relevant functions.
% FSMs
SCRUPLE~\cite{Paul1992,Paul1994} uses a finite automata-based comparison of a code query and code examples.
After turning both into a finite automata, a code pattern automaton interpreter compares two pieces of code and reports a match if the automaton reaches the final state.

\subsubsection{Solver-based Matching}
\label{sec:ir solver}

% SMT solver
Code search engines that represent the behavior of code in the form of constraints often use SMT solvers to match queries against code examples~\cite{Stolee2016,Jiang2018}.
The indexing phase in this case consists of a static analysis that extracts constraints describing input-output relationships.
Then, the retrieval phase checks with an SMT solver whether the constraints of a code example satisfy the input-output examples that a user provides as the query.
The idea was first proposed by \citet{Stolee2016} and later refined and generalized by \citet{Jiang2018}.

%% MP: Removing sequential search as a separate 'IR technique', because 
%% (a) it actually isn't (but avoids indexing and retrieval) and
%% (b) some of the above techniques are also sequential, I think (e.g., solver-based matching)

%\subsubsection{Sequential Search}
%\label{sec:ir sequential}
%
%% sequential search based on pair-wise comparison
%
%There are also techniques based on a pairwise comparison between queries and code examples.
%\citet{Inoue2020} performs a sequential pattern matching algorithm between two token sequences for the query and the target~\cite{Gusfield1997}.
%% [seems sequential] structural information around the code (both query code and searched code)
%\citet{Holmes2005} use different kinds of heuristics. \emph{Inheritance heuristic} based on methods that have the same super classes, \emph{call heuristics} based on the same method calls in the body of the function, and \emph{use heuristics} based on the types used in the method. 
%
%\cite{Lawall2016} Prequel

% ontology
%\cite{Durao2008}: augment a key word-based code search with (i) classification of code into domains and (ii) an ontology of software-specific words; both help refine the search results

\begin{finding}
	The most used approaches for indexing and retrieval are feature vector-based retrieval and, more recently, deep learning-based models. The first approach needs less data and represents query and source code both as interpretable feature vectors. The second approach needs more data for training a model, e.g., to embed both queries and code source into a joint vector space.
\end{finding}

\subsection{Granularity of Retrieved Source Code}
\label{sec:ir granularity}

Different code search engines retrieve code at different levels of granularity.
We categorize the existing approaches into four kinds of granularity.
First, many search engines retrieve \emph{code snippets}, which may range from a single line of code to multiple consecutive lines that implements a specific task.
Second, other search engines focus on the \emph{method}-level, i.e., these approaches retrieve entire methods.
Third, users can also search at the \emph{class}-level, where code search engines return entire classes.
Finally, there also are search engines that operate at the \emph{application or library}-level, which we do not cover in full detail here.
Table~\ref{tab:granularity} summarizes the approaches and the granularity they use.
The same approach may appear in multiple rows~\cite{Ossher2009, Linstead2009} if it supports multiple kinds of granularity.

The design decision of the granularity level to target is very important for a code search engine, because it affects what a user can search for.
For example, snippets of code-level can be useful to search for code that provides an example of how to use an API~\cite{Bajracharya2010}.
The disadvantage of retrieving code snippets is that they may be incomplete and thus hard to directly reuse.
Searching at the method-level can be useful for finding full methods that already solve a specific task~\cite{Poshyvanyk2009}, which a user may directly reuse.
Finally, class-level and application or library-level approaches are useful to find entire components to reuse. Due to the more coarse-grained granularity, the number of suitable results may be limited though.

\begin{table}[]
	\centering
	\caption{Granularity of source code extracted by code search approaches.}
	\label{tab:granularity}
	\small
	\begin{tabular}{@{}lp{33em}@{}}
		\toprule
		Granularity           & Approaches \\ \midrule
		Snippet of code       &
		\cite{Li2016, Balachandran2015, Gu2018, Luan2019,Mishne2012,Bajracharya2010, Paul1992, Paul1994,Diamantopoulos2018,Luan2019,Sachdev2018,David2014a,Chen2001,Linstead2009,Chatterjee2009,Ossher2009, Chen2018,Fujiwara2019,Kisub2018,Zhou2018,Holmes2005,Kisub2018,Mukherjee2020, Sun2022, Chai2022,	Sirres2018, Nie2016,Ye2020, Premtoon2020, Lu2018, Vinayakarao2017,Wightman2012,Stolee2014, Salza2022}    \\
		Method                & \cite{Poshyvanyk2009,Lv2015,Jiang2018,Linstead2009,Ossher2009,Husain2019,Nguyen2016,Mcmillan2013,Wu2019,Sivaraman2019, Podgurski1993, Stolee2016, Nguyen2017,Reiss2009,Lu2015,Lemos2011}     \\
		Class                 & \cite{Linstead2009,Ossher2009,Reiss2009}     \\
		Application or library & \cite{Bajracharya2006,Linstead2009,Ossher2009,Akbar2019}      \\ \bottomrule
	\end{tabular}%
\end{table}

%% file: sections/rankingpruning.tex
\section{Ranking and Pruning of Search Results}
\label{sec:rankingpruning}

After retrieving code examples that likely match a query, many code search engines rank and prune the results before showing them to the user.
This step is critical to enable users to quickly see the most relevant matches.
In the following, we discuss and compare different ranking (Section~\ref{sec:ranking}) and pruning (Section~\ref{sec:pruning}) approaches.

\subsection{Ranking of Search Results}
\label{sec:ranking}

%We summarize the main approaches in Table~\ref{tab:ranking_distance}.

%\begin{table}[]
%	\centering
%	\caption{Main ranking approaches.}
%	\label{tab:ranking_distance}
%	\begin{tabular}{@{}llll@{}}
%		\toprule
%		Cosine Similarity & Euclidean Distance & Hamming Distance & Custom Distance \\ \midrule
%		\cite{Sachdev2018,Chen2001,Chen2018,Cambronero2019,Takuya2011,Nguyen2017}               & \cite{Lee2011,Balachandran2015,Lee2010}                 & \cite{Bajracharya2010}               & \cite{David2014a,Lu2015}               \\ \bottomrule
%	\end{tabular}%
%
%\end{table}

\subsubsection{Standard Distance Measures}

The by far most common ranking approach is to rely on a distance measure implicitly provided by the retrieval component of a code search engine (Section~\ref{sec:indexretrieve}).
In this approach, the query and each code example are first represented as feature vectors, then a standard distance measure gives the distance between a query vector and a code vector, and finally code examples with a smaller distance to the query are ranked higher.
For example, this ranking approach can be implemented using cosine similarity~\cite{Sachdev2018,Chen2001,Cambronero2019,Takuya2011,Chen2018,Ling2021, Sun2022, Salza2022},
Hamming distance~\cite{Bajracharya2010}, and
Euclidean distance~\cite{Balachandran2015,Lee2010,Lee2011}.

\subsubsection{Custom Ranking Techniques}

In addition or as an alternative to standard distance measures, several search engines rely on custom ranking techniques.
% custom
\citet{David2014a} propose a variation of string edit distance to compute the similarity between two sequences of assembly instructions. The basic idea is to treat each instruction as a letter and to use a table that provides a heuristic distance between assembly instructions.
Another approach~\cite{Li2016} ranks code examples using two scores that are based on the number of tokens that match the given natural language description and the length of a code snippet, respectively. 
\citet{Sachdev2018} augment the rank obtained via cosine similarity with custom rules, such as the number of query tokens present in the candidate, to re-rank the list of results.
Another example is from~\citet{Lu2015}. They compute a representative set of words for each method and then rank results via a normalized intersection of these words.
COSAL~\cite{Mathew2021} combines multiple custom ranking techniques, which compare two pieces of code based on their token similarity, structural similarity, and behavioral similarity, respectively.

% based on code that a developer has already written
Some ranking approaches look beyond the given query by also considering the code a developer is editing while making a query.
For example, when building a query vector, \citet{Takuya2011} give more weight to occurrences of tokens near the cursor position, to find programs that contain similar fragments to the code around the cursor position.
In a similar vein, \citet{Wightman2012} uses features of the programmer's source code to rank and filter prospective snippet results, including variable types and names, the cursor position within the abstract syntax tree, and code dependencies.
A higher rank here means that a code example uses more of the existing variables etc., and hence will require fewer modifications.

% learning-based
Some more recent ranking approaches are based on machine learning models.
For example, the Lancer approach~\cite{Zhou2019} fine-tunes a pre-trained BERT model\footnote{https://github.com/huggingface/pytorch-pretrained-BERT} to predict whether a code example matches the given, incomplete method, and then ranks code examples based on the predicted score.
\citet{Ye2020} compute the similarity score using two parameters retrieved with a code summarization model and a code generation model, based on a dual learning technique. 

\begin{finding}
	To rank search results, engines often use standard algorithms, such as cosine similarity and Euclidean distance, or they implement custom variations of these techniques. 
\end{finding}

\subsection{Pruning of Search Results}
\label{sec:pruning}

Orthogonal to ranking, several search engines also prune search results that are unlikely to be of interest to the user.
The most straightforward pruning technique is to discard results based on similarity threshold.
For example, some approaches discard all candidates with a similarity lower than some threshold~\cite{Jiang2018, Chatterjee2009}, while others show only the top N results in the output~\cite{Kisub2018,Rahman2018,Lu2018}.
Another way of pruning search results is to merge similar code examples, assuming that a user likely wants to see only one of them.
For example, \citet{Mishne2012} merge similar method call paths relevant to the query to remove redundancy in the final results. 
Aroma~\cite{Luan2019} uses a greedy algorithm based on parse tree comparison to find and remove redundant code snippets, followed by re-ranking the pruned search results.

\begin{finding}
	Filtering by a threshold and merging similar results are the most popular techniques for pruning code search results.
\end{finding}

%% file: sections/study.tex
\section{Empirical Studies of Code Search}
\label{sec:studies}

\begin{table}
  \caption{Overview of empirical studies on code search.}
  \label{tab:studies}
  \small
  \setlength{\tabcolsep}{9pt}
  \centering
  \begin{tabular}{@{}l|lllllllllllll@{}}
\toprule
& \begin{sideways}\begin{minipage}{7em}\citet{Singer1997}\end{minipage}\end{sideways}
& 
\begin{sideways}\begin{minipage}{8em}\raggedright\citet{Sim1998}\end{minipage}\end{sideways}
& \begin{sideways}\begin{minipage}{8em}\raggedright\citet{Ko2006}\end{minipage}\end{sideways}
& \begin{sideways}\begin{minipage}{8em}\raggedright\citet{Sim2011}\end{minipage}\end{sideways}
& \begin{sideways}\begin{minipage}{8em}\raggedright\citet{Panchenko2011}\end{minipage}\end{sideways}
& \begin{sideways}\begin{minipage}{8em}\raggedright\citet{Bajracharya2012}\end{minipage}\end{sideways}
& \begin{sideways}\begin{minipage}{8em}\raggedright\citet{Sadowski2015}\end{minipage}\end{sideways}
& \begin{sideways}\begin{minipage}{8em}\raggedright\citet{Rahman2018a}\end{minipage}\end{sideways}
\\
\midrule
Topic of study: \\
\hspace{1em} Usage of development tools
&\yes&&\yes&&&&& \\
\hspace{1em} Usage of search tools
&&\yes&&\yes&\yes&\yes&\yes&\yes \\
\hspace{1em} Activities of developers
&\yes&\yes&\yes&&&&\yes& \\

Methodology: \\
\hspace{1em} Questionnaire
&\yes&\yes&\yes&&&&\yes& \\
\hspace{1em} Log analysis
&\yes&&&\yes&\yes&\yes&\yes&\yes \\
\hspace{1em} Observing developers
&\yes&&\yes&&&&& \\

Searched code: \\
\hspace{1em} Single project
&\yes&\yes&\yes&&&&& \\
\hspace{1em} Multiple projects within organization
&\yes&&&&\yes&&\yes& \\
\hspace{1em} Many open-source projects
&&&&\yes&&\yes&&\yes \\

\bottomrule
  \end{tabular}
\end{table}

The wide adoption of code search in practice raises various interesting questions about the way developers search for code.
This section discusses empirical studies related to how, when, and why developers search for code and what tools they use for this purpose.
We include all such empirical studies that we are aware of and that fit the selection criteria given in Section~\ref{sec:intro}.
We start by describing the experimental setups used in these studies (Section~\ref{sec:studies setups}) and then present some of their main results (Section~\ref{sec:studies results}).
Table~\ref{tab:studies} gives an overview of the discussed studies, including the topics they address, the methodologies they use, and the amount of code searched through by the studied developers.\\

\subsection{Setups of Empirical Studies}
\label{sec:studies setups}

While practically all empirical studies address the broad questions of how, when, and why developers search for code, they use different setups and methodologies to address this question.
Early studies~\cite{Singer1997,Ko2006} are mostly about what activities developers spend their time on and what tools they use, including tools used for code search.
In contrast, more recent studies~\cite{Sim1998,Sim2011,Panchenko2011,Bajracharya2012,Sadowski2015,Rahman2018a} focus specifically on code search tools and what activities they are used for.

We see three kinds of methodologies, and sometimes combinations of them: questionnaires answered by developers~\cite{Singer1997,Sim1998,Ko2006}, analyses of logs of search engines~\cite{Singer1997,Sim2011,Panchenko2011,Bajracharya2012,Sadowski2015,Rahman2018a}, and observing developers, e.g., by shadowing them~\cite{Singer1997} or by recording their screens~\cite{Ko2006}.
The first two kinds of studies are typically based on data gathered from tens~\cite{Sim1998,Ko2006,Sim2011} to hundreds~\cite{Sadowski2015} of developers.
In contrast, log analysis often covers much larger datasets, ranging between tens of thousands~\cite{Panchenko2011} and ten million~\cite{Bajracharya2012} logged activities.

The studies also vary by the amount of code that is searched through by the studied developers.
Reflecting the general trends in code search engines, early studies are about searching through a single project~\cite{Singer1997,Sim1998,Ko2006}, whereas later studies are about searching through multiple projects, either within a larger organization~\cite{Panchenko2011,Sadowski2015} or the open-source ecosystem~\cite{Sim2011,Bajracharya2012,Rahman2018a}.

\subsection{Results of Studies and their Implications}
\label{sec:studies results}

A recurring finding in studies is that code search is among the most common activities developers spend their time on.
Early studies report that \emph{grep}, \emph{find}, and its variants are used on a regular basis~\cite{Singer1997,Sim1998}.
For example, measurements of tool invocations by \citet{Singer1997} shows that \emph{grep} and its variants are the second-most frequently used developer tools, right after the compiler.
The observational study by \citet{Ko2006} also reports code search to be a common activity.
However, their definition of ``searching for  code'' only partially matches ours, because we assume that there is an explicitly formulated query, whereas they also mean reading code to find a specific code location.
A study at Google based on a specialized code search engine shows that the average developer is involved in five search sessions per day, with a total of twelve daily queries~\cite{Sadowski2015}.
Overall, these findings highlight the importance of code search in software development, motivating researchers and practitioners to work on code search techniques.

Several studies investigate the goals that developers have when searching for code.
The three most commonly reported goals are
finding example code to reuse, e.g., when trying to understanding how to use an API (between 15\%~\cite{Sim1998} and 34\%~\cite{Sadowski2015} of all searches),
program understanding (between 14\%~\cite{Sim1998} and 29\%~\cite{Sadowski2015} of all searches), and 
understanding and fixing a bug (between 10\%~\cite{Sadowski2015} and 20\%~\cite{Sim1998} of all searches).
Beyond these three goals, a long tail of other goals is reported, such as understanding the impact of a planned code change, finding locations relevant for a code clean-up, understanding the coding style used within an organization, and identifying the person responsible for a particular piece of code.
A perhaps surprising finding is that developers also often use code search as a quick way to navigate through code they are already familiar with~\cite{Sadowski2015}.

Being a central element of every search, queries and their properties have received some attention in studies.
A study of the Koders code search engine finds most queries to be short, with 79\% of users providing only a single search term~\cite{Bajracharya2012}.
In contrast, other studies report longer queries, e.g., an average of 4.2 terms per query in a study across five search engines~\cite{Sim2011}, and of 4.7 terms for code-related queries given to Google's general-purpose search engine~\cite{Rahman2018a}.
Beyond the size of queries, several studies investigate what terms are used in queries.
\citet{Bajracharya2012} find that both code queries and natural language queries are common.
Comparing code-related queries with general-purpose web search queries, \citet{Rahman2018a} find that code queries use a smaller vocabulary.
Another interesting finding related to queries is that they are frequently reformulated within a search session~\cite{Sim2011,Bajracharya2012,Sadowski2015}, even more often than general web search queries~\cite{Rahman2018a}.

Finally, some studies analyze and compare how effective code search engines are at providing useful search results.
One study reports that between 25\% and 60\% of all queries are effective, depending on the kind of query, where ``effective'' means that the search results cause the user to download a relevant piece of code~\cite{Bajracharya2012}.
Another study compares specialized code search engines with general-purpose web search engines.
It finds that the former are more effective when searching for entire subsystems, e.g., a library to use, whereas the latter are more effective for finding individual blocks of code~\cite{Sim2011}.
The same study also reports that it is easier to find reference examples than components that are reusable as-is.

\begin{finding}
Empirical studies of developers show that they commonly perform code search to reach various goals, including code understanding, finding code to reuse, and quickly navigating to code the developer already knows.
\end{finding}

%% file: sections/future.tex
\section{Open Challenges and Research Directions}
\label{sec:future}

\subsection{Support for Additional Usage Scenarios}

Each code search engine focuses on one or more usage scenarios, such as finding examples of how to use a specific API or finding again some code that a developer has previously worked on.
In addition to the currently supported usage scenarios, we envision future work to support other search-related developer tasks.
For example, developers may want to search not only through a static snapshot of code, but also search for specific kinds of changes in the version histories of projects.
Searching for changes could help developers, e.g., to understand how a particular API usage typically evolves, to find examples of code changes similar to a change a developer is currently working on, or to find code changes that have introduced bugs.
\citet{Lawall2016} and Di~\citet{DiffSearch_arXiv_2022} propose promising first steps into this direction.
Another example of a currently unsupported usage scenario is cross-language search.
In this scenario, a user formulates a code query in one programming language to find related code written in  another programming language.
Such cross-language search could help developers transfer their knowledge across languages, e.g., when a developer knows how to implement a particular functionality in one but not in another language.

\subsection{Cross-Fertilization with Code Completion and Clone Detection}

Code search relates to other problems that have received significant attention by researchers and that offer opportunities for cross-fertilization.
One such problem is code completion, i.e., the problem of suggesting suitable code snippets while the developer is writing code in an integrated development environment (IDE).
Recent large-scale language models used for code completion offer a functionality similar to code search.
For example, a typical usage scenario of GitHub's Copilot tool\footnote{\url{https://github.com/features/copilot}} and the underlying Codex model~\cite{Chen2021} takes a short natural language description of a desired functionality and maps it to a code snippet offering that functionality.
This usage scenario is closely related to code search engines that receive free-form queries (Section~\ref{sec:free-form queries}).
It remains an open challenge to apply successful techniques from code search in code completion, and vice versa.
Another problem that is strongly related to code search is clone detection~\cite{roy2007survey}.
Similar to code search engines that accept programming language queries (Section~\ref{sec:PL-based queries}), clone detectors try to find code that is similar to a given code example.
A key difference is that clone detection tries to find multiple code examples that implement the same functionality (possibly with syntactic and semantic differences that do not affect the overall behavior), whereas code search tries to retrieve code that offers more functionality than the given query.
Despite these different goals, there is potential for cross-fertilization of the two related fields, e.g., by adapting effective representations of code (Section~\ref{sec:indexretrieve}) or mechanisms for pruning search results (Section~\ref{sec:pruning}).

\subsection{Learning-Based Code Search}

Given the tremendous progress in machine learning, adopting the newest models to code search is likely to offer new opportunities to code search.
In particular, we identify three open challenges.
First, future work could benefit from the increasingly effective code representation models proposed in the neural software analysis field~\cite{NeuralSoftwareAnalysis} to compute a vector representation of code and of queries, which can then be used to identify code examples similar to a given query.
Some instances of this idea have already been presented~\cite{Gu2018}, but as code representation models keep increasing, adopting new models is likely to also improve code search.
Second, future work could design models that not only retrieve code, but also generalizing examples seen during training into new code that fits a query.
Open challenges here include to formulate code search as a zero-shot learning problem~\cite{Brown2020} and to adapt models that combine question answering with retrieval~\cite{lee2019latent}.

%, improve the ranking loss for code retrieval~\cite{Feng2020,Ye2020} or build a large dataset of human written query and related code snippet~\cite{Ling2021}.
%Moreover, to index and retrieve code examples, most existing approaches focus on one representation of the code.
%Future work would explore combinations of different representations, e.g., by comparing a query and code examples based on both the static code structure and the runtime behavior of the code, as pioneered by \citet{Mathew2021}.
%Another example is to improve graph representation and comparison of text and code to improve semantic code search~\cite{Ling2021}.

\subsection{Deployment and Adoption in Practice}

Code search is an area of strong interest by academic researchers, tool builders in industry,  and practitioners.
Despite the already impressive use of code search by developers, we see several open challenges related to its deployment and adoption in practice.
On the one hand, there are challenges faced by people who are running and maintaining a code search engine.
For example, the problem of how to incrementally re-index a code corpus when the code is evolving has not yet received significant attention by researchers.
A naive approach is to continuously re-index the entire corpus, which is likely to unnecessarily repeat significant computational effort.

On the other hand, there are challenges faced by users of code search engines.
While various techniques have been proposed for the core components of code search, its user interface is receiving less attention, with some noteworthy exceptions, such as some of the query expansion techniques discussed in Section~\ref{sec:querypre}.
An interesting line of future work could be to automatically formulate clarification questions, e.g., in natural language, which could allow a user to prune the search space with a single click.
Another promising direction is to support users in defining code queries (Section~\ref{sec:query}) by adding automatic code completion features known from IDEs into the interface of a search engine.
Finally, future work could design ``query-less'' search engines that suggest suitable code snippets while a developer is writing code, without the need to explicitly formulate a query.
First steps toward that goal have been taken, e.g., by \citet{Brandt2010} and \citet{Takuya2011}.

\subsection{Common Datasets and Benchmarks}

An effective way to foster further progress in the research field is to offer reusable datasets for evaluating and comparing code search engines.
Ideally, such a dataset should be realistic, large-scale, and cover multiple programming languages.
Several benchmark datasets have been proposed and are use by parts of the existing work.
One kind of benchmarks consists of groups of semantically equivalent implementations, e.g. BigCloneBench~\cite{DBLP:conf/icsm/SvajlenkoIKRM14}, Google Code Jam\footnote{\url{https://codingcompetitions.withgoogle.com/codejam}}, and AtCoder\footnote{\url{https://atcoder.jp/}}.
Such benchmarks are particularly useful to evaluate code-to-code search engines (Section~\ref{sec:PL-based queries}), as one implementation in a group can be used as a query, while the other implementations are expected to show up among the results.
Benchmarks that come with executable test cases, e.g., those derived from coding competitions, are also useful to evaluate approaches based on dynamic analysis (Section~\ref{sec:runtime behavior}).

Another kind of benchmarks offers pairs of natural language queries and code.
For example, the CodeSearchNet challenge offers such a dataset, which has been automatically gathered and covers Go, Java, JavaScript, PHP, Python, and Ruby code~\cite{Husain2019}.
In a similar vein, CodeXGLUE offers query-code pairs for Python and Java~\cite{Lu2021}.
The Search4Code dataset provides code-related queries extracted from Bing search queries via a weakly supervised discriminative model~\cite{Rao2021}.
Instead of relying on automated extraction of datasets, CoSQA is a benchmark of pairs of natural language queries and code examples that have been manually annotated~\cite{Huang2021}.

We anticipate future work to build even more than today upon these datasets, either as a benchmark to evaluate a novel code search engine, or as a training dataset to learn from.
There also are opportunities for creating datasets and benchmarks that go beyond those available today.
For example, the community would benefit from a dataset that not only includes queries and search results, but also information on how developers act on search results, e.g., by selecting lower-ranked results or by copying and adapting code examples.
An interesting challenge for benchmarks used to evaluate learning-based code search is how to ensure that a model does not see the benchmark during learning.
As large-scale, pre-trained models~\cite{Feng2020,Guo2021,Chen2021}, which often are trained on a large fraction of all publicly available source code, are becoming increasingly popular, the chances that a publicly available benchmark is coincidentally used during training increases.

%% file: sections/conclusion.tex
\section{Concluding Remarks}

This article provides a comprehensive overview of 30 years of research on code search.
Given the huge amounts of existing code, searching for specific code examples is a common activity during software development.
To support developers during this activity, various techniques for finding relevant code have been proposed, with an increase of interest during recent years.
We discuss what kinds of queries code search engines support, and give an overview of the main components used to retrieve suitable code examples.
In particular, the article discusses techniques to pre-process and expand queries, approaches toward indexing and retrieving code, and ways of pruning and ranking search results.
Our article enables readers to obtain an overview of the field, or to fill in gaps of their knowledge of the state-of-the-art.
Based on our survey of past work, we conclude that code search has evolved into a mature research field, with solid results that have already made an impact on real-world software development.
Despite all advances, many open challenges remain to be addressed in the future, and we hope our article will provide a useful starting point for addressing them.

\section*{Acknowledgment}
This work was supported by the European Research Council (ERC, grant agreement 851895), and by the German Research Foundation within the ConcSys and DeMoCo projects.